\documentclass[12pt]{article}

\usepackage{fullpage}
\usepackage{cite}
\usepackage{graphicx}
\usepackage{amsmath}
\usepackage{amssymb}
\usepackage{array} 
\usepackage{slashed} 

\allowdisplaybreaks
\title{}
\date{}

\def\beq{\begin{equation}}
\def\eeq{\end{equation}}
\def\beqa{\begin{eqnarray}}
\def\eeqa{\end{eqnarray}}

\newcommand{\be}{\begin{equation}}
\newcommand{\ee}{\end{equation}}
\newcommand{\bea}{\begin{eqnarray}}
\newcommand{\eea}{\end{eqnarray}}

\def\slash#1{#1 \hskip-0.45em /}

\begin{document}
\bibliographystyle{utphys}


\titlepage

\begin{flushright}
Nikhef/2015-033\\
ITF-UU-15/14
\end{flushright}

\vspace*{1.2cm}

\begin{center}
{\Large \bf Top physics at the LHC}

\vspace*{1cm} \textsc{
  Vittorio del Duca$^{a,b}$ and
  Eric Laenen$^{c,d,e}$ } \\

\vspace*{1.5cm}

$^a$ Institute for Theoretical Physics, ETH Z\"urich, 8093 Z\"urich, Switzerland

\vspace*{0.3cm} 

$^b$ INFN, Laboratori Nazionali di Frascati, 00044 Frascati (Roma), Italy  

\vspace*{0.3cm} 

$^c$ Nikhef, Science Park 105, NL-1098 XG Amsterdam, The Netherlands

\vspace*{0.3cm} 

$^d$ ITFA, University of Amsterdam, Science Park 904, Amsterdam, The Netherlands

\vspace*{0.3cm} 

$^e$ ITF, Utrecht University, Leuvenlaan 4, Utrecht, The Netherlands

\vspace*{1cm}

\textit{Invited review for International Journal of Modern Physics A}

\end{center}

\vspace*{1cm}

\begin{abstract}
\noindent
We review the present situation in top quark physics, in these early days
of Run II of the LHC. We take mostly a Standard Model 
perspective, showing recent results, and review the
key concepts and results of the associated theoretical predictions. 
The issues we discuss are the top quark mass, 
top quark pair and single top production, production in
association with other particles, charge asymmetry and top quark decay.
\end{abstract}


\section{Introduction}

Ever since its discovery\cite{Abe:1995hr,Abachi:1995iq} 
in 1995 by the CDF and D0 experiments at the Tevatron, the top 
quark has been in or near the center of attention in high-energy phyiscs. Its 
remarkably large mass, still the largest of any known elementary particle, 
implies that it couples strongly to the agents of electroweak symmetry breaking, 
making it both an object of interest itself, and a tool to investigate that
mechanism in detail. 

The history of heavy flavours anyway is such 
each of them has taught us much about Nature. From the charm quark we learned
that the Standard Model is consistent, through the GIM \cite{Glashow:1970gm}
mechanism. Moreover, its discovery cemented the belief in QCD as the
quantum theory of the strong interactions. From the bottom quark we
learned that a complete third family was there to find, in turn
allowing for weak CP violation \cite{Kobayashi:1973fv} 
to be part of the Standard Model. However,
although already discovered 20 years ago, the top quark has not \emph{yet} taught
us fundamentally new insights. The top may well do this
in the coming decade after all, a belief that rests on top's attributes.

The top quark couples to other particles 
through various (chiral, vector, scalar) structures according to the Standard
Model Lagrangian. In our search for physics beyond the Standard Model,
all of these bear scrutiny for deviations, and there is therefore
much to test.  Such precise scrutiny is feasible because the large top mass
implies that hadronization effects do not occur and spin information is preserved.

With the Tevatron having made the first precious thousands top quarks, 
leading to its discovery and tests of some of its properties,
the LHC is a genuine top quark factory, in particular in Run II, which is now 
underway. The data gathered already and especially upcoming data 
will allow us to study the top quark and its behavior in LHC collisions
in great detail, if also the theoretical descriptions and simulations
are of commensurate quality. 

Here we provide a compact review of some of the key aspects of top quark physics,
largely from a Standard Model point of view. 
We highlight key issues from a mostly conceptual standpoint, and list
the present state of affairs in terms of calculations and
corresponding experimental analyses. 
We refer to other excellent reviews\cite{Bernreuther:2008ju,Han:2008xb,Quadt:2007jk,Schilling:2012dx,JABEEN:2013mva} for 
more extensive explanations. In section \ref{sec:top-prop-decays} we discuss mostly
issues regarding properties of the top itself, and characteristics of its decay.
In section \ref{sec:top-production} we mostly discuss its production, either in
pairs, singly, or in association. We end with a brief conclusion.

\section{Top properties and decays}
\label{sec:top-prop-decays}

In this first section we give a brief description of how the top quark is embedded in the Standard
Model, and motivates physics beyond it. We also discuss aspects of its
decay and properties such as mass and spin.

\subsection{Top in the Standard Model}

We recall the various interactions of the top quark field $t(x)$ 
in the Standard Model Lagrangian. The interaction with gluons is a vectorlike coupling involving
an $SU(3)$ generator in the fundamental representation
\begin{equation}
  \label{eq:1}  
g_s \bar{t}_i(x) \gamma^\mu \left[T^a\right]_{ij} t_j(x) G_\mu^a(x)\,,
\end{equation}
where $i,j$ label QCD colour charge. 
The interaction with photons is also simply vectorlike and proportional
to the top quark electric charge
\begin{equation}
  \label{eq:2}
  \frac{2}{3}e\,\, \bar{t}(x) \gamma^\mu   t(x) A_\mu(x)\,.
\end{equation}
Its charged weak interaction is left-handed and flavour-changing
\begin{equation}
  \label{eq:3}
  \frac{g_w}{2\sqrt{2}} V_{tf}\,\, \bar{t}(x) \gamma^\mu(1-\gamma_5) f(x) W_\mu(x),
\quad f={d,s,b}\,,
\end{equation}
while its neutral weak interaction is flavour-conserving  and parity violating
\begin{equation}
  \label{eq:16}
  \frac{g_w}{4\cos\theta_W} \, \bar{t}(x) \gamma^\mu\left((1-\frac{8}{3}\sin^2\theta_W) -\gamma_5\right) t(x) Z_\mu(x)\,.
\end{equation}
Finally, the interaction of the top with the Higgs boson is of the Yukawa type
\begin{equation}
\label{eq:23}
  y_t\,h(x)\bar{t}(x) t(x)\,,
\end{equation}
with a coupling constant $y_t = \sqrt{2}m_t/v$ directly proportional
to the top quark mass $m_t$, and $v$ is the Higgs vacuum expectation value. 

Beyond these, effective interactions such as for flavour-changing neutral 
currents, occur due to loop corrections. They can be calculated and are
generally very small compared to the ones above. All these interactions, either
elementary or effective, could be modified in structure and strength 
by virtual effects due to new interactions associated with physics 
beyond the Standard Model. This is a particularly interesting line of investigation 
for the top quark, if only because it evidently has a large coupling to the electroweak
symmetry breaking sector (the Yukawa coupling $y_t$ in Eq.~(\ref{eq:23}) is
almost exactly 1 in the Standard Model).  It is then important to test
these structures in detail, and indeed this is the thrust behind the field of top physics. 

\subsection{Top beyond the Standard Model}
\label{sec:beyond}

Driving most motivations for physics beyond the Standard Model is the 
fact that the Higgs mass seems unnaturally small. The top quark features prominently in this 
argument as the main culprit for creating this situation.
When considering Standard Model one-loop corrections to the inverse
Higgs boson propagator there are contributions from the $W$ and $Z$ bosons, the Higgs boson itself,
and, most importantly, the top quark. Using an ultraviolet cut-off regulator $\Lambda$ they 
can be added to the bare Higgs mass squared $m_{H,B}^2$ to form the renormalized 
Higgs mass $m_H$
\begin{multline}
  \label{eq:6}
  m_H^2 = m_{H,B}^2 + \left(-\frac{3}{8\pi^2} y_t^2 \right) \Lambda^2 \; [\mathrm{top}]\;
+ \left(\frac{9}{64\pi^2} g^2  \right) \Lambda^2 \;
[\mathrm{vector\,\, bosons}]\; \\
+ \left(\frac{1}{64\pi^2} \lambda^2  \right) \Lambda^2 \; [\mathrm{Higgs}]\,.
\end{multline}
Because symmetry is not enhanced by setting
the Higgs mass to zero, renormalization is not necessarily
multiplicative \cite{'tHooft:1979bh}, and the divergent corrections 
are in fact quadratic in $\Lambda$. 
Eq.~(\ref{eq:6}) shows that when $\Lambda$ is of order, say, the GUT
scale, cancellations to many digits are required among these contributions, which
seems a very fine-tuned setup. 

Being the main troublemaker, the top may in fact also point to possible
new physics in which this finetuning is avoided. A popular model is
supersymmetry where \emph{stop} quark loops naturally provide
the cancellations that finetuning does in the Standard Model. But also 
in supersymmetry phenomenology the top quark 
plays an important role: if it weren't for the top quark (and stop squark)
corrections to the lightest Higgs boson mass, the Minimal Supersymmetric
Standard Model (MSSM) would predict the latter to be lighter than the $Z$ boson,
and the MSSM would have been ruled out already. The maximum viable mass
for the Higgs mass is thus about $140$ GeV, comfortably above the
measured value of $125$ GeV.

Top could play an even more central role in the Higgs mechanism, in
that its dominant contribution to the running of a Higgs potential
parameter down from the GUT scale in fact leads to a negative
eigenvalue for the Higgs mass matrix, thereby even explaining electroweak
symmetry breaking \cite{Ibanez:1982fr}. 

We also note that in the last few years the precise value of the top mass has been
moving further into the spotlight due to its role in regards to the 
stability of the electroweak vacuum\cite{Branchina:2013jra}; the current 
value suggests that the vacuum is meta-stable\cite{Bezrukov:2012sa,Degrassi:2012ry,Buttazzo:2013uya,Alekhin:2012py}. 

In short, there is good reason to study the top quark in detail, 
what its properties are, how it is produced, and how it decays. We begin 
with the latter.

\subsection{Top quark decay}
\label{sec:top-quark-decay}

\def\abs#1{\left|#1\right|}
\def\bentarrow{\:\raisebox{1.3ex}{\rlap{$\vert$}}\!\longrightarrow}
\def\HW{{\small HERWIG}}
\def\DiracSlc#1{\slash\!\!\!#1}
\def\DiracSld#1{\slash\!\!\!\!#1}
\newcommand\sss{\scriptscriptstyle\rm}
\newcommand\bl{\overline{l}}
\newcommand\bt{\overline{t}}
\newcommand\bn{\overline{\nu}}
\newcommand\bu{\overline{u}}
\newcommand\bv{\overline{v}}
\newcommand\mV{m_{V}}
\newcommand\GammaV{\Gamma_{V}}
\newcommand\mVi{m_{V_i}}
\newcommand\GammaVi{\Gamma_{V_i}}
\newcommand\mW{m_{W}}
\newcommand\GammaW{\Gamma_{W}}
\newcommand\mt{m}
\newcommand\Gammat{\Gamma}
\newcommand\vecVcp{V_{Vl}}
\newcommand\axlVcp{A_{Vl}}
\newcommand\vecVicp{V_{V_il_i}}
\newcommand\axlVicp{A_{V_il_i}}
\newcommand\vecWcp{V_{Wl}}
\newcommand\axlWcp{A_{Wl}}
\newcommand\vecZcp{V_{Zl}}
\newcommand\axlZcp{A_{Zl}}
\newcommand\vecGcp{V_{\gamma l}}
\newcommand\axlGcp{A_{\gamma l}}
\newcommand\gw{g_{\sss W}}
\newcommand\thw{\theta_{\sss W}}
\newcommand\veps{\varepsilon}
\newcommand\Vtb{V_{tb}}
\newcommand\mydot{\!\cdot\!}
\newcommand\tM{\tilde{M}}
\newcommand\tB{\tilde{B}}
\newcommand\tV{\tilde{V}}
\newcommand\clH{{\mathbb H}}
\newcommand\clS{{\mathbb S}}
\newcommand\EVprjmap{{\cal P}_{\clH\to\clS}}
\newcommand\bSigma{\overline{\Sigma}}
\newcommand\xMCB{\Big|_{\sss {\rm MC}}}
\newcommand\Gfun{{\cal G}}
\newcommand\np{n^\prime}
\newcommand\npp{n^{\prime\prime}}
\newcommand\hnp{\hat{n}^\prime}
\newcommand\Ms{M}
\newcommand\Ns{N}
\newcommand\As{A}
\newcommand\pt{p_{\sss T}}
\newcommand\kt{k_{\sss T}}

\noindent The top quark decay characteristics play, directly or indirectly, an important role in
studying the top quark at colliders. The top quark 
width is largely due to decays to a $W$-boson and a 
bottom quark. But because the top quark mass is
much larger than the sum of the $W$ and $b$ masses, the width 
is sufficiently large to pre-empt top quark hadronization. The rapid
decay of the top quark moreover enables transmision of top quark spin
information to final states, giving us an important tool to test
the role of top quark spin. At the same time, the width-to-mass ratio
$\Gamma/m$ of the top quark is small enough that, for many purposes, the notion
of top quark as a stable particle makes sense. This is effectively implemented through the 
narrow width approximation (NWA), which factorizes the production and 
decay processes. But, although the NWA works well for
many, especially inclusive observables, it is still necessary to test its quality
well, given how carefully we aim to study the top quark's behavior.

The top width itself
is very difficult to determine in a hadron collider, 
though a recent experimental inference of the width in the context of single top
$t$-channel production was performed by D0\cite{Abazov:2012vd} finding
$\Gamma = 2.00^{+0.47}_{-0.43}$~GeV, and CDF \cite{Aaltonen:2013kna}, finding
$1.10 < \Gamma < 4.05$~GeV at the 68\% confidence level. 
An optimal determination would require a threshold scan for pair production 
at a $e^+e^-$ collider.

The NWA full separation of production and decay is indeed an approximation, and 
there are corrections to it. Besides the intrinsic uncertainty of
order $\Gamma/m$, there are also non-factorizable corrections
from virtual partons that connect production and decay amplitudes. Another
irreducible class of corrections is from diagrams with the same final
state but having no intermediate top quark.

Let us briefly describe how the NWA works for the decay process of the top quark 
\begin{equation}
  \label{eq:21}
t(p)\;\longrightarrow\; W^+(q)+b(r)
\;\longrightarrow\; l^+(k_1)+\nu(k_2)+b(r)\,,
\end{equation}
where the top has been produced in the production process
\begin{equation}
a(P_1)+b(P_2)\,\longrightarrow\,
t(p)+X(x)\,.
\label{stopproc}
\end{equation}
We shall also see how spin correlations can be included in the NWA.
The squared amplitude for the combined process reads
{\renewcommand\jot{8pt}
\begin{eqnarray}
\abs{A}^2&=&\frac{\gw^4\abs{\Vtb}^2}{64}
\frac{1}{(p^2-\mt^2)^2+(\mt\Gammat)^2}
\frac{1}{(q^2-\mW^2)^2+(\mW\GammaW)^2}
\nonumber\\*&\times&
\bu(r)\gamma^\mu(1-\gamma_5)\left(\DiracSlc{p}+\mt\right)MM^*\gamma^0
\left(\DiracSlc{p}+\mt\right)(1+\gamma_5)\gamma^\rho u(r)
\nonumber\\*&\times&
\bu(k_2)\gamma_\mu(1-\gamma_5)v(k_1)
\bv(k_1)(1+\gamma_5)\gamma_\rho u(k_2)\,,
\label{stopampsq}
\end{eqnarray}}
\noindent where the top and $W$ propagators (and their widths) are shown on the
first line, while the other two lines contain the squared matrix
element for (off-shell) top production, and (off-shell) $W$ decay.
Here $M=\Gamma u(K)$, with $\Gamma$ a combination of $\gamma$-matrices,
and $k_1, k_2$ the four-momentum of a fermion entering the hard scattering.

The narrow top width approximation $\Gammat\to 0$ now amounts to 
making the replacement
\begin{equation}
\frac{1}{(p^2-\mt^2)^2+(\mt\Gammat)^2}\;\longrightarrow\;
\frac{\pi}{\mt\Gammat}\,\delta\left(p^2-\mt^2\right)\,,
\label{topnw}
\end{equation}
yielding an on-shell condition for the top quark momentum $p$.
Summing over spins one may now write the squared amplitude
in (\ref{stopampsq}) as 
\begin{equation}
  \label{eq:22}
\sum_{spin}\abs{A}^2=\frac{\pi}{\mt\Gammat}\sum_{\lambda\lambda^\prime}
\tM_\lambda \rho_{\lambda\lambda^\prime}
\tM^*_{\lambda^\prime}\,
\delta\left(p^2-\mt^2\right)\,.
\end{equation}
The matrix $\rho$ is the decay spin-density matrix, encoding spin correlations
between production and decay, with $\lambda, \lambda'$ labelling the top quark
spin states. The above procedure to include 
spin correlations in the NWA can be implemented in Monte Carlo programs, even
in those matched to NLO \cite{Frixione:2007zp,Artoisenet:2012st,Motylinski:2009kt}. This works
in many cases very well. Other studies in this regard  for $t\bar{t}$ production
can be found in Refs.~[\cite{Melnikov:2009dn}] and [\cite{Denner:2012yc}]. 

One should however not take the validity of the NWA for granted for 
all observables. Especially for those cases where there is a sizeable 
contribution from intermediate top quarks that are not near their mass
shell this is an important issue. In these phase space regions there 
can moreover be appreciable contributions from subprocesses producing the same
final state, but having no intermediate top quark. 
A recent study \cite{Papanastasiou:2013dta}
investigated the off-shell effects in $t$-channel single-top production, in part as
a test of the NWA. Also an effective theory approach
\cite{Falgari:2011qa} was compared to the exact calculation,
including non-resonant diagrams and off-shell effects in the
aMC@NLO \cite{Alwall:2014hca} framework using the complex mass scheme 
\cite{Denner:2006ic,Denner:2005fg,Denner:1999gp}.
It was shown that indeed the NWA approximation does
not always work well, in particular for observables sensitive to the $W$-$b$ invariant mass,
while the effective theory approach does track the exact NLO calculation rather well.

Another NLO study \cite{Heinrich:2013qaa} comparing $t\bar{t}$
production plus decay in the NWA with $W^+W^-b\bar{b}$ production, the
latter including also singly resonant and non-resonant contributions,
found that these contributions have a signifcant impact on shapes of
distributions, and thereby also the uncertainty of top mass
measurements. 

For most of the results discussed below, however, except where stated otherwise,
the NWA is taken, and should be a good approximation.

\subsection{The Higgs-top interaction and the $W$ polarization}
\label{sec:higgs-top-inter}

Top interacts with the Higgs boson through the Yukawa interaction 
\begin{equation}
  \label{eq:4}
  y_t\, h\bar{t} t\,,
\end{equation}
where $y_t = \sqrt{2}m_t/v$. This is a relation that can be kept at the
renormalized level as well, whatever the choice of renormalization
scheme for the top quark mass (about which more below).
As remarked, for the top quark pole mass of about 173~GeV and with $v=246$ GeV, 
$y_t=1$ to a very good approximation.

It is interesting to note that the large Yukawa coupling $y_t$
of the top with the Higgs boson is related to the large fraction
of top quarks decaying into longitudinal $W$ bosons.
In fact, although it is reasonable to expect that in the decay (\ref{eq:21})
of a top quark to a $W$ boson, $t\to W^+ + b$, the width be proportional
to the weak coupling $g^2$ and to the top mass, a direct 
calculation~\cite{Kane:1991bg} shows that the expression for the width reads
\begin{equation}
\Gamma(t\to W^+ b) \propto g^2\, m_t\, a\,
\left( 1 + {\cal O}\left( a^{-1}\right) \right)\,,
\end{equation}
with
\begin{equation}
a = \frac{m_t^2}{2m_W^2} = \frac{y_t^2}{g^2}\,.
\end{equation}
Note that the width is enhanced by a factor $a$ (about 2.3) with respect to the naive expectation.
Looking at the breakdown of this result to different intermediate $W$ polarizations, we see
that the decay to transversely polarized $W$ bosons is in line with the naive expectation,
while the $a$ enhancement is due to the longitudinal polarization of the $W$ 
bosons~\cite{Kane:1991bg}. In fact, the Goldstone boson equivalence
theorem \cite{Chanowitz:1985hj},
which states that the longitudinally polarized $W$ boson acts as a
Goldstone boson (a member of the Standard Model Higgs doublet), predicts
that in the limit $m_t \gg m_W$ the
width of the top decaying into a longitudinal $W$ boson behaves as\cite{Peskin:1995ev}
\begin{equation}
\Gamma(t\to W_L^+ b) \propto g^2\, m_t\, a\,.
\end{equation}
We can then write that the fraction of longitudinally polarized $W$ bosons is approximately given by
\begin{equation}
{\cal F}_L = \frac{\Gamma(t\to W_L^+ b)}{\Gamma(t\to W^+ b)} \approx \frac{a}{1+a}\,.
\end{equation}
Because the ratio $a$ of the squared top and $W$ masses, or equivalently of
the top Yukawa and the gauge couplings, is about 2.3, we expect that about 70\%
of the $W$ bosons are longitudinally polarized.
In fact, a precise computation\cite{Czarnecki:2010gb} which includes the NNLO QCD corrections as well the
leading electroweak contributions yields ${\cal F}_L = 0.687(5)$. This value is well within the experimental error bands,
since an early combination of CDF and D0 Tevatron Run II data\cite{Aaltonen:2012rz} yielded ${\cal F}_L = 0682 \pm 0.057$,
while CDF\cite{Aaltonen:2012lua} obtained ${\cal F}_L = 0.726 \pm 0.094$ using the full set of Tevatron Run II data.
At the LHC, CMS has found ${\cal F}_L = 0.682\pm 0.045$ in the 7 TeV run\cite{Chatrchyan:2013jna}, and
${\cal F}_L = 0.720 \pm 0.054$ in the 8 TeV run\cite{Khachatryan:2014vma}.

\subsection{Top mass}
\label{sec:top-mass}

The top quark property that is perhaps most central in many aspects of top physics
is its mass. We already mentioned its role in the issue of stability of the
Higgs potential. 
From Run I and Run II data and for an integrated luminosity of up to $9.7\,\mathrm{fb^{-1}}$, 
the Tevatron experiments~\cite{Tevatron:2014cka} have measured the mass with a total uncertainty of 
0.64~$\mathrm{GeV}/c^2$, {\it i.e.} to an accuracy of less than 0.4\%.
From the run at a centre-of-mass energy of 7~TeV and for an integrated luminosity of up 
to $4.9\,\mathrm{fb^{-1}}$, the LHC experiments~\cite{ATLASCMStopmasscomb} 
have measured the mass with a total uncertainty of 0.95~$\mathrm{GeV}/c^2$,
\begin{align}
  \mathrm{CDF/D0}\, : & \quad 174.34 \pm 0.37 \mathrm{(stat)} \pm 0.52 \mathrm{(sys)}\, \mathrm{GeV}/c^2 
  \,,\nonumber \\
  \mathrm{ATLAS/CMS}\, : & \quad 173.29 \pm 0.23 \mathrm{(stat)} \pm 0.92 \mathrm{(sys)}\,  \mathrm{GeV}/c^2\,. 
\end{align}

The Tevatron data from Run II at a centre-of-mass energy of 1.96~TeV and for an integrated luminosity of up to $8.7\,\mathrm{fb^{-1}}$
have been combined with the LHC data mentioned above. The resulting worldwide combination is
\begin{equation}
\mathrm{ATLAS/CDF/CMS/D0}\, :  173.34 \pm 0.27 \mathrm{(stat)} \pm 0.71 \mathrm{(sys)}\,  \mathrm{GeV}/c^2\,,
\end{equation}
with a total uncertainty of 0.76~$\mathrm{GeV}/c^2$.

CMS~\cite{Khachatryan:2015hba} has also provided a combination of LHC data from the run at a centre-of-mass energy of 7~TeV 
and for an integrated luminosity of up to $5.1\,\mathrm{fb^{-1}}$, with the data from the run at a centre-of-mass energy of 8~TeV 
and for an integrated luminosity of up to $19.7\,\mathrm{fb^{-1}}$, measuring the mass with
a total uncertainty of 0.49~$\mathrm{GeV}/c^2$, {\it i.e.} to an accuracy of less than 0.3\%,
\begin{equation}
\mathrm{CMS}\, : 172.44 \pm 0.13 \mathrm{(stat)} \pm 0.47 \mathrm{(sys)}\,  \mathrm{GeV}/c^2\,. 
\end{equation}

Together with an accurately measured $W$ boson mass, a precisely known
top mass severely constrains the mass range of the Higgs boson~\cite{Baak:2012kk}.
Indeed the measured Higgs boson mass seems quite consistent given present accuracies.
Therefore its precise measurement is of considerable importance, and
therefore also its careful definition. This is necessary because for the
top, being coloured and thus subject to confinement, defining the mass
is indeed subtle. 

A natural definition of an elementary particle mass is based on the location of the pole of the full 
propagator, i.e. the pole mass. 
After summing self-energy
corrections the full quark propagator reads
\begin{equation}
\label{eq:11}
\frac{1}{\slashed{p} - m_0 - \Sigma(p,m_0)}\,,
\end{equation}
where $\Sigma$ contains  $1/\epsilon$ UV divergences from loop integrals.
Renormalization (here at one loop) now amounts to replacing the bare mass $m_0$ by an
expression involving the renormalized mass $m$
\begin{equation}
\label{eq:15}
  m_0 = m \Big(1 + \frac{\alpha_s}{\pi}\Big[\frac{1}{\epsilon} + z_{finite} \Big]  \Big)\,,
\end{equation}
after which the UV divergences cancel in (\ref{eq:11}). The choice of $z_{finite}$
determines the scheme. Choosing it such that 
\begin{equation}
\label{eq:19}
   \frac{1}{\slashed{p} - m_0 - \Sigma(p,m_0)}
= \frac{c}{\slashed{p} - m}
\end{equation}
is the pole-mass scheme, which amounts to pretending that the particle
can be free and long-lived. However, because 
no quark can ever propagate out to infinite times due to confinement,
such a pole only exists in perturbation
theory, and its location is intrinsically ambiguous by $\mathcal{O}(\Lambda_{QCD})$
\cite{Beneke:1994sw,Bigi:1994em,Smith:1997xz}. 

Experimentally, the mass of the top quark is most often reconstructed 
by collecting the jets and leptons from its decay.  The decay channels used are the dilepton channel - two isolated leptons
with opposite charge and at least two jets\cite{Chatrchyan:2012ea,ATLAStopmassdilepton}; 
the lepton~$+$~jets channel - an isolated lepton and at least four jets\cite{ATLAS:2012aj,Chatrchyan:2012cz}; 
the all-hadronic channel\cite{Chatrchyan:2013xza,ATLAStopmassalllepton}. 
However, soft particles originating from both within and outside these jets may affect the reconstructed
mass. Moreover, various experimental methods used (e.g. track quality
cuts) and corrections do not have a clean perturbation theory description. 
Though it is considered generally a measurement of the pole mass, the full
procedure has led to some discussion about what the precise ``scheme''
is of the mass thus measured, and to the possibility
of considering a Monte Carlo mass, which would track closely but not
be 
quite the same as the pole mass\cite{Hoang:2008xm}.

Although the experiments in this way reconstruct the pole mass (or something close to it),
theoretically it would be more desirable to have a short-distance mass, free
of $\mathcal{O}(\Lambda_{QCD})$ ambiguities.
Such is the $\overline{\mathrm{MS}}$ mass
$\bar{m}(\mu)$, evaluated at some scale $\mu$, whose
relation to the pole mass is known in QCD to three loops analytically\cite{Melnikov:2000qh}
and four loops numerically\cite{Marquard:2015qpa}. For $\mu$
one often takes the implicit value found when intersecting the $\bar{m}(\mu)$
curve with the $\bar{m}(\mu)=\mu$ axis, yielding $\bar{m}(\bar{m})$.
The $\overline{\mathrm{MS}}$ mass
$\bar{m}(\mu)$ may be extracted 
indirectly, by comparing, for instance, the measured inclusive cross
section with the theoretical one expressed in the  $\overline{\mathrm{MS}}$
mass \cite{Langenfeld:2009wd}. Of course, such an indirect measurement will depend upon
the accuracy of the theoretical calculation of the inclusive cross
section, and its sensitivity to the mass.
A recent evaluation by D0 \cite{Abazov:2011pta} along these
lines yields $\bar{m}(m) = 154.5\, \mathrm{GeV}/c^2$, if the cross section is evaluated to NLO~$+$~NNLL
accuracy, \cite{Ahrens:2010zv} or $\bar{m}(m) = 160.0\, \mathrm{GeV}/c^2$, if the cross section is evaluated to NNLO accuracy,
\cite{Moch:2008qy,Langenfeld:2009wd} with an uncertainty in both cases of about 5 $\mathrm{GeV}/c^2$.

There are other definitions of short-distance masses, inspired by the
top quark pair production near threshold at a future $e^+e^-$
collider.  One is the $1S$~mass\cite{Beneke:1999qg, Hoang:1999zc,Hoang:2000yr}, which is related to the peak position in the
cross section for $e^+e^-\to t\bar{t}$, and is defined as half the perturbative mass of a fictitious toponium
ground state, where the top quark is assumed to be stable. The relation between the $1S$~mass and the pole mass
is known to higher orders in QCD. As both the $1S$~mass and the $\overline{\mathrm{MS}}$ mass are short-distance masses,
the relation between them is $\mathcal{O}(\Lambda^2_{QCD}/m)$\cite{Hoang:1999zc}.
At lepton colliders, it might be feasible to determine the $1S$ top
mass with a precision of about 100~MeV\cite{Juste:2013dsa}  (see this
reference also for an overview of other methods).
Another promising short-distance mass is the potential-subtracted mass \cite{Beneke:1998rk}, which
employs the fact that the IR sensitive part of the pole, discussed in
section \ref{sec:top-quark-decay} cancels against the IR sensitivity
of the top-antitop Coulomb potential in threshold production. 

Current procedures to measure the top mass are the {\it template method}, which uses distributions of top mass values 
obtained from the event kinematics, and compares them to distribution templates for reference top mass values, and the
{\it matrix-element method} (MEM) \cite{Kondo:1988yd,Kondo:1991dw,Dalitz:1991wa}, which uses the (tree level) matrix elements 
to estimate the likelihood of each experimental event for kinematic configurations which come from events of a given top mass.
Improvements of the methods above have been proposed. As regards the former, the template overlap method for infrared safe
jet observables \cite{Almeida:2010pa} has been put forward, which is based on the fact that the energy flow in jets which come 
from the decay of highly boosted top decay products is different from the one in jets which come from the QCD background.
As regards the MEM, the inclusion of QCD radiation effects \cite{Alwall:2010cq} and the computation of NLO weighted events
\cite{Campbell:2012cz} have been proposed.

Alternative methods to measure the top mass are also under
consideration. Mostly use proxy variables that are to a varying degree 
sensitive to the mass and can be accurately calculated and measured.
One of them uses the leptonic final states of a $J/\psi$
\cite{Kharchilava:1999yj}: one may consider the process $pp\to (t\to W^+ + b\to W^+ + J/\psi) + (\bar{t}\to W^- + \bar{b})$
and require that the $W^-$ decays hadronically, the $W^+$ decays leptonically and the $J/\psi$ decays into leptons, typically
muons. Then the invariant mass distribution of the $J/\psi$ and an isolated lepton
can be used to evaluate the top mass. Since no jets are involved, the measurement is not plagued by jet energy scale (JES)
uncertainties, which allows for an accurate reconstruction of the $m_{J/\psi l}$ invariant mass, with a projected 
${\cal O}(1 {\rm GeV})$ error on the top mass. However, the leptonic decays reduce the rate substantially,
and a large integrated luminosity, of the order of tens of ${\rm fb}^{-1}$, is required. The error on the top mass evaluation
can be further improved by including the NLO QCD corrections to production and decay \cite{Biswas:2010sa}.

Some methods use the kinematic distributions of the dilepton channel
to either determine the pole mass while being little sensitive to long-distance effects~\cite{Frixione:2014ala}
or to perform a simultaneous evaluation of the top-quark, $W$-boson and neutrino masses, basing it
on end-point determinations in the kinematic distributions~\cite{Chatrchyan:2013boa}. 
This may be convenient in the investigation of 
New Physics models, where several masses in a decay chain may be unknown.

Other methods use the invariant mass $m_{t\bar{t}}$ of the $t\bar{t}$ pair \cite{Frederix:2007gi}, or 
examine $t\bar{t}$ production in association with a jet \cite{Alioli:2012hj,Alioli:2013mxa}, and use the invariant mass 
$m_{t\bar{t} jet}$ of the $t\bar{t}$-jet system. These methods complement top mass 
measurements from the $t\bar{t}$ total cross section \cite{Langenfeld:2009wd,Beneke:2011mq,Ahrens:2011px}.

Finally, a novel method exploits the large top Yukawa coupling to extract the top mass from loop effects in flavour physics 
observables~\cite{Giudice:2015toa}.

\subsection{Spin and  angular correlations}
\label{sec:top-simulation}

Part of the attractiveness of the top quark as a study object is
its power to self-analyze its spin, through its purely left-handed
SM weak decay. This is both a useful aid in signal-background separation, 
and itself a property worthy of detailed scrutiny, as certain new physics models
could introduce right-handed couplings. The correlation between top spin and
directional emission probability for its decay products is expressed
through
\begin{equation}
  \label{eq:13}
  \frac{d\ln \Gamma_f}{d\cos\chi_f} = \frac{1}{2}\left(1+\alpha_f \cos\chi_f  \right)\,,
\end{equation}
where $|\alpha_f| \leq 1$, with 1 indicating 100\% correlation.
Note that in the NWA the correlation between the production process and the spin of the
produced top quark is indicated in eq.~(\ref{eq:22}).
For the dominant decay mode
\begin{equation}
  \label{eq:14}
  t \rightarrow b + W^+ (\rightarrow l^+ + \nu)\,,
\end{equation}
at lowest order, we have $\alpha_b = -0.4, \alpha_\nu=-0.3, \alpha_W = 0.4, \alpha_l=1$. 
QCD corrections to these values are small\cite{Czarnecki:1990pe,Brandenburg:2002xr}.
The charged lepton direction (or the down-type
quark in a hadronic decay of the intermediate $W$) is indeed nearly 100\% correlated with the
top quark spin. This is notably more than for its parent $W$ boson, a consequence of interference
of two amplitudes with different intermediate $W$ polarizations. 

In single-top quark production, which occurs via the charged weak
interaction, the top is produced left-handed, so 
a correlation should be a clear feature of the production process
and serve as a discriminant to suppress the background. 
In top quark pair production a correlation of an individual quark with a fixed direction 
is absent\footnote{There is a tiny correlation due $Z$-boson mediated
  production.}, however there is a clear correlation between the top and anti-top spins. The size of the correlation depends on the choice
of reference axes $\hat{\mathbf{a}},\hat{\mathbf{b}}$ \cite{Bernreuther:2000yn,Bernreuther:2001bx,Mahlon:1997uc}.
At the Tevatron the
beam direction $\hat{\mathbf{a}}=\hat{\mathbf{b}}=\hat{\mathbf{p}}$ is 
good choice, at the LHC the helicity axes 
$\hat{\mathbf{a}}=\hat{\mathbf{b}}=\mathbf{\hat{k}_{top}}$ should give
near-maximal correlation 
\begin{equation}
  \label{eq:12}
  \frac{d\sigma}{d\cos\theta_a d\cos\theta_b}
 = \frac{\sigma}{4}\left(1 + B_1 \cos \theta_a + B_2 \cos \theta_b
 - C \cos\theta_a \cos\theta_b
 \right)\,.
\end{equation}
Indeed, the correlation coefficient $C$ depends on the correlation
axis. Thus, at LO in QCD, the values for 
$\{C_{hel},C_{beam}\}$ at the Tevatron (LHC) is 
$\{0.47,0.93 \}$ ($\{0.32,-0.01\}$). NLO corrections modify these
numbers somewhat \cite{Bernreuther:2004jv}. BSM models
that influence the pair production mechanism (e.g. new resonances)
could noticeably influence these correlations.

There is also the interesting possibility of azimuthal
angular distributions as indicators of new physics. Thus, in the
dilepton decay channel, after an invariant mass cut, $t\bar{t}$
spin correlations may be revealed through the $\Delta\phi$
distribution of leptons in the laboratory frame \cite{Mahlon:2010gw}. This
observable is quite robust, as the 
correlation remains visible even after summing over spurious neutrino
momentum resolutions, and persists at NLO
\cite{Melnikov:2011ai}.

Other angular distributions can function as quite selective probes of 
new physics \cite{Godbole:2010kr,Choudhury:2010cd}. 
For instance, if a $Z'$ would polarize tops at production, the
azimuthal asymmetry
\begin{equation}
  \label{eq:10}
  A_\phi = \frac{\sigma(\cos \phi_l>0) -\sigma(\cos \phi_l>0) }{\sigma(\cos \phi_l>0) +\sigma(\cos \phi_l>0) }\,,
\end{equation}
where $\phi_l$ is the azimuthal angle of the lepton with respect to
the beam-top plane, would be sensitive to the amount of left-handed and
right-handed coupling, even more so when judicious cuts on the $p_T$ of the
top are chosen. When a charged Higgs is present, such an
asymmetry  would help distinguish \cite{Huitu:2010ad,Godbole:2011vw}
$Wt$ from $H^-t$ production.

\section{Top production}
\label{sec:top-production}

Having discussed issues concerning top quark decays, we now turn to
aspects of top quark production.
 In this section we discuss a number of much studied top quark production 
observables. For each we review the theoretical issues, and present experimental status. 

\subsection{Top pair production cross section}
\label{sec:pair-prod-cross}

Let us first discuss the cross section measurements from the four experiments that have
collected tops in large quantities. Note that besides cross sections 
inferred from specific final states,
combinations are being made that consist of analyses with different final
states, with somewhat different integrated luminosities. 

At the Tevatron at 1.96 TeV the measured pair production cross sections, based on 
almost all of the collected data, are
\begin{align}
 \mathrm{CDF}: &\;   7.63 \pm 0.31 \, (\mathrm{stat}) \pm 0.39 \,
(\mathrm{sys})\pm 0.15\, (\mathrm{th}) \; \mathrm{pb}\,, \nonumber \\
 \mathrm{D0}:  &\;  7.56 \pm 0.20 \; (\mathrm{stat})\; \pm 0.56\,
  (\mathrm{sys}) \; \mathrm{pb}\,, \nonumber \\
 \mathrm{Tevatron \; combined}: &\;  7.60 \pm 0.20 \; (\mathrm{stat})\;
  \pm 0.36\, (\mathrm{sys}) \; \mathrm{pb} \,.
\end{align}
The combination shown \cite{Aaltonen:2013wca} has a measured uncertainty of
about 5.4\%.  The best present calculation \cite{Czakon:2013goa}
yields $7.24+0.23-0.27$pb (3.4\%). 

The measured pair production cross sections by ATLAS and CMS  at 7 TeV are
\begin{align}
 \mathrm{ATLAS}: &\;   177 \pm 3 \,(\mathrm{stat}) \pm \begin{array}{c}+8\\-7\end{array} \,
(\mathrm{sys})\pm 7\, (\mathrm{lum})\; \mathrm{pb} \nonumber \,,\\
 \mathrm{CMS}:&\;    166 \pm 2 \,(\mathrm{stat}) \pm 11 \,
(\mathrm{sys})\pm 8\, (\mathrm{lum}) \; \mathrm{pb}\,,\nonumber \\
\mathrm{LHC\; Combined}: &\;  173.3 \pm 2.3 \,(\mathrm{stat}) \pm 
9.8 \, (\mathrm{sys}) \; \mathrm{pb} \,.
\end{align}
The combined result \cite{ATLAS:2012dpa,CMS:2013sca} and its measured uncertainty of 
about 5\% is to be compared to the best present calculation \cite{Czakon:2013goa} which
yields $172+6.4-7.5$pb (5.7\%). 

For the 8 TeV data we quote two recent results, for the ATLAS
di-lepton ($e\mu$) for 20.3pb$^{-1}$, and the CMS di-lepton ($e\mu$) channel for
5.3pb$^{-1}$, respectively
\begin{align}
  \mathrm{ATLAS }: &\;  242.4 \pm 1.7 \,(\mathrm{stat}) \pm 5.5 \,
(\mathrm{sys})\pm 7.5\, (\mathrm{lum}) \; \mathrm{pb}\nonumber \,,\\
\mathrm{CMS }: &\;  239.0 \pm 2.6 \,(\mathrm{stat}) \pm 11.9 \,
(\mathrm{sys}) \pm 6.2\, (\mathrm{lum}) \; \mathrm{pb}\,, \nonumber \\
\mathrm{LHC\; Combined  }: &\;  241.5 \pm 1.4 \,(\mathrm{stat}) \pm 
5.7 \, (\mathrm{sys}) \pm 6.2\, (\mathrm{lum}) \; \mathrm{pb} \,,
\end{align}
with an uncertainty of about 3.5\% \cite{ATLAS:2014xxx,CMS:2014gta}. The
best current calculation \cite{Czakon:2013goa} yields
$245.8+8.8-10.6$pb (5.6\%). Interestingly, the experimental
uncertainty is now again smaller than the theoretical one, providing a
challenge to theory.

First results at 13 TeV are now appearing, with both CMS
\cite{Khachatryan:2015uqb} and ATLAS
results, still with large errors, in agreement with theory predictions.

For both colliders and for each collision energy the measurements 
are clearly in agreement with each other, and 
with the best theoretical calculations, which we
discuss below. The remarkable agreement among different
collision types and energies gives us solid
confidence in the value and structure of the top quark QCD
coupling.

Let us now review the status of, and main ideas behind theoretical
calculations for top quark pair production. The inclusive top 
pair production cross section has always played a role that is both useful and
instructive in perturbative QCD, because it only involves QCD
couplings. It moreover features
a truly large produced mass whose effects play a crucial role in 
both in the matrix elements and the phase space measure.
The NLO corrections were
computed\cite{Nason:1987xz,Nason:1989zy,Beenakker:1989bq,Beenakker:1991ma}
 in the late 80's.
For many years these were among the most difficult one-loop
calculations done. In these first calculations phase space was (partially) integrated over
in analytical way; a fully differential calculation 
was completed shortly thereafter \cite{Mangano:1991jk}. 
The combination of such a fully differential calculation with parton
showers, such as \texttt{MC@NLO}
\cite{Frixione:2002ik,Frixione:2003ei} and
\texttt{POWHEG} \cite{Nason:2004rx,Frixione:2007vw} is now the state
of art at this order in perturbation theory. These codes
combine the virtues of the exclusiveness of a parton shower event
generator with the accuracy of a NLO calculation. 

A recent major development has been the completion of the full NNLO
calculation\cite{Baernreuther:2012ws,Czakon:2012zr,Czakon:2012pz,Czakon:2013goa} 
for the inclusive pair production cross section.
This is indeed a milestone in top quark physics, even in
perturbative QCD as a whole. The result is a hadronic cross section
computed with a theoretical accuracy at the few percent level, as already mentioned.  The
calculations require NNLO corrections to both the $q\bar{q}$ and the $gg$
channel, as well as the NLO corrections to the
$qg$ channels.  For both the $q\bar{q}$ and $gg$ channel, the second
order corrections are composed of three classes of contributions, some
computed at different times by various authors.  These are (i) the
two-loop corrections, (ii) the one-loop plus one real emission
correction, and (iii) the double real emission contribution.  The
double-real emission calculations were computed earlier
\cite{Czakon:2011ve,Abelof:2011jv,Bernreuther:2011jt}.  The one-loop,
one real emission contributions are known, since the NLO calculation
for $t\bar{t} + \, \mathrm{jet}$ is available
\cite{Dittmaier:2007wz,Melnikov:2010iu}.  The two-loop virtual
corrections have been performed\cite{Czakon:2007ej,Czakon:2007wk,Bernreuther:2004ih,Bonciani:2010mn,Baernreuther:2013caa}.
The methods used so far are a combination of analytical and numerical
ones. The latter involve solving differential equations in the
kinematic invariants, which requires a highly accurate initial
condition (chosen to be at high energy), and avoiding singularities in
the equations. The double-real emission contribution was achieved
through the use of a method called \texttt{STRIPPER} \cite{Czakon:2011ve}. The
one-loop, one-real emission diagrams could be computed with
well-established techniques.  

\begin{figure}
  \centerline{\includegraphics[width=0.7\columnwidth]{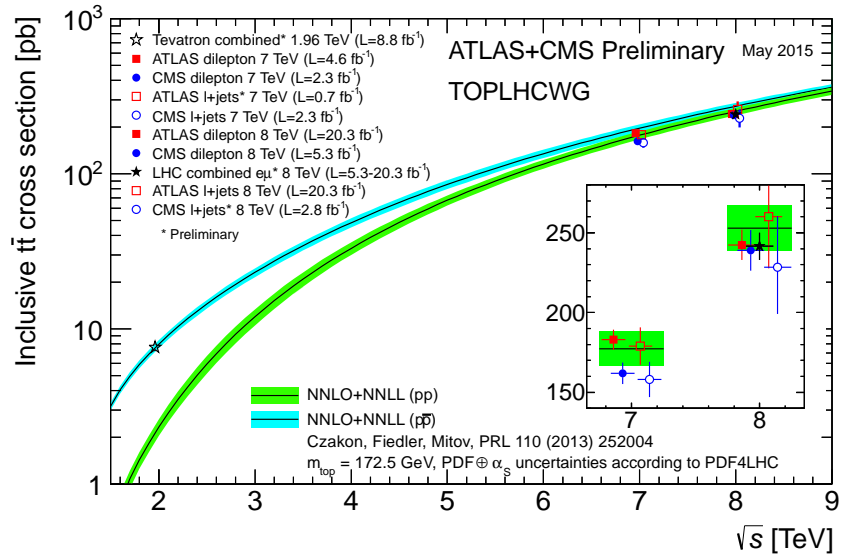}}
  \caption{Data and theory for the inclusive top quark pair production cross 
section at Tevatro and LHC as function of the
    collider c.m. energy, compared to recent measurements of ATLAS
    and CMS, compiled by the Top Physics LHC Working Group.} \label{fig:sigmatot}
\end{figure}

The full calculation, altogether a major tour-de-force, 
has good perturbative convergence and very
small uncertainties. Given these properties and the excellent
agreement with measurements, as shown in Fig.~\ref{fig:sigmatot}, 
a comparison of theory and data for 
the inclusive cross section can be used more prosaically to infer useful knowledge about
the gluon density. A first study in this direction was done
\cite{Czakon:2013tha}, demonstrating the feasibility and desirability of this.
The top cross section has now been included in the NNPDF3.0 global fit \cite{Ball:2014uwa}.
Electroweak corrections to top pair production have also been computed
\cite{Bernreuther:2006vg,Kuhn:2006vh,Moretti:2006nf}, which can be
large in certain phase space regions, depending on transverse
momentum. They can also impact the charge asymmetry \cite{Kuhn:2011ri}. 
Calculations including off-shell
effects are beginning to appear as well \cite{Falgari:2011qa,Bevilacqua:2010qb}. 

On top of the exactly calculated orders one can add arbitrarily high orders in
approximately using threshold resummation. 
The latter also underlies some theoretical estimates of the top
quark charge asymmetry, discussed in section
\ref{sec:charge-asymmetry}, as well as various distributions,
so let us review this method briefly here.

When the top quark pair is produced near threshold in hadronic collisions, logarithms
whose argument represents the distance to threshold in the perturbative
series become numerically large. 
The definition of the threshold depends on the observable. Thus, 
for the inclusive cross section the threshold is given by the condition
$T_1:\, s-4m^2=0$. For the transverse momentum distribution we have
$T_2:\, s-4(m^2+p_T^2) = 0 $, and for the doubly differential distribution in 
$p_T$ and rapidity we can choose 
\begin{equation}
\label{eq:20}
T_3:\, s-4(m^2+p_T^2)\cosh y = 0  \quad \mathrm{or} \quad  T_3:\, s + t + u -2m^2 = 0\,.
\end{equation}
The perturbative
series for any of these (differential) cross sections can be in
general expressed as
\begin{equation}
  \label{eq:7}
  d_i \sigma (T_i) = \sum_n \sum_k^{2n} \alpha_s^{n} c^i_{n,k} \ln^k (T_i)\,,
\end{equation}
plus non-logarithmic terms.
Here $T_i$ represents any of the threshold conditions, suitably
normalized, for the observables
enumerated by $i$. Note that it is allowed to use e.g. $T_2$ for the
inclusive cross section, by first analyzing $d\sigma/dp_T$ and then integrating
over $p_T$, and similarly for $T_3$. For any complete fixed order calculation this will give the same 
answer, but if one only selects the logarithmic terms because the exact answer
is unknown, numerical differences will occur. Such kinematic
differences can then be viewed as theoretical uncertainties \cite{Kidonakis:2001nj}. 

The threshold logarithms result from integration over phase space
regions where the emitted gluons are soft and/or collinear to their on-shell emitter.
Resummation concerns itself with carrying out the sum in
Eq.~(\ref{eq:7}), and the result takes the generic form
\begin{equation}
  \label{eq:9}
   d\sigma = \exp \left(Lg_0(\alpha_sL) + g_1(\alpha_sL) + \alpha_sg_2(\alpha_sL)
+\ldots \right)\times C(\alpha_s)\,.
\end{equation}
Including up to the function $g_i$ in the exponent amounts to N$^i$LL
resummation, with the coefficient $C(\alpha_s)$ then evaluated
to order $i-1$. Key benefits of threshold resummation are (i) gaining all-order control of the
large terms which plague
fixed-order perturbation theory, to restore predictive
power, and (ii) reduction of scale uncertainty. Regarding the first
point, the reason these resummable terms are large for the top quark
pair inclusive cross section is that, while
the hadronic cross section is Sudakov suppressed near threshold, 
the PDF's are over-suppressed, which the
partonic cross section must then partially compensate for.
Regarding the second point, when examining the 
sources of scale dependence, they occur both in the PDF and in 
the partonic cross section now both \emph{in the exponent}, which
improves the cancellation\cite{Sterman:2000pt}.

The state-of-the-art accuracy for threshold resummation for 
inclusive pair production cross section at present is NNLL
\cite{Czakon:2009zw,Ahrens:2011mw,Kidonakis:2009ev,Kidonakis:2010dk}. A consistent combination of 
NNLL accuracy in both threshold and Coulomb corrections has now also
been achieved \cite{Beneke:2011mq}. The latter are 
only relevant for threshold $T_1$ and behave as $(\alpha_s/\beta)^n$, with
$\beta^2 = 1-4m^2/s$. From such all-order results, approximate NNLO results were 
constructed before the completion of the exact calculation. 
This is of particular interest for thresholds $T_1$ and $T_3$. 
The latter, being dependent on $t$ and $u$,  then allows estimating 
threshold resummation corrections to the forward-backward
asymmetry, a point we return to in section
\ref{sec:charge-asymmetry}. 
Other approximate NNLO calculations use threshold $T_3$ and results
based on these \cite{Kidonakis:2010dk} are typically larger than for $T_1$, closer to
the exact result; estimates are also made for approximate NNNLO\cite{Kidonakis:2014isa}.
As mentioned above, calculations using $T_3$ can 
assign ambiguities due to using either pair-invariant mass (PIM) or
one-particle inclusive (1PI) kinematics in the precise definition of
the threshold to a theoretical error \cite{Kidonakis:2001nj,Ahrens:2011px}.

The various theoretical calculations are available in a number of
codes, such as \texttt{HATHOR}\cite{Aliev:2010zk}
(contains full NNLO corrections, and possibility of using a running
top quark mass), \texttt{TOP++} \cite{Czakon:2011xx}
(contains full NNLO corrections, and NNLL threshold resummation),
\texttt{TOPIXS }\cite{Beneke:2012wb} (contains NLO, approximations
for NNLO, and NNLL resummation, including Coulomb corrections). 

In the above, the top quarks are treated as on-shell stable
particles, using the narrow-width approximation. 
It is interesting to include in the full description also
the top quark decays, including the effects of off-shellness and
spin-correlations. Thus, one considers then as final state of interest
$WWbb$. For zero $b$-quark mass two groups have computed the NLO corrections 
to this production process \cite{Denner:2010jp,Bevilacqua:2010qb},
establishing an interesting tool to study such effects.

\subsection{Charge asymmetry}
\label{sec:charge-asymmetry}

A different test of the QCD production mechanism of top quarks, one that has received
much attention in recent years, is the charge asymmetry: the normalized difference in 
production rate between top and anti-top at some fixed angle or rapidity
\begin{equation}
\label{eq:17}
A_t(y) = \frac{N_t(y)-N_{\bar{t}}(y)}{N_t(y)+N_{\bar{t}}(y)}\,.
\end{equation}
While electroweak production via a $Z$-boson could produce a (very small) asymmetry at LO, QCD itself
produces it at $\mathcal{O}(\alpha_s^3)$ through a term proportional
to the SU(3) $d_{abc}$ 
symbol \cite{Nason:1989zy,Beenakker:1991ma,Kuhn:1998kw,Dittmaier:2007wz}.
A more precise look \cite{Kuhn:1998kw} shows that the asymmetry is due
to an interference between C-odd and C-even terms. In top quark
pair production in the $q\bar{q}$ channel this amounts to the Born
diagram and the one-loop box diagram, respectively.
When computing such an interference contribution, the asymmetry reveals
itself in terms of the Mandelstam variables $t$ and $u$ as terms that are odd under
$t \leftrightarrow u$ interchange, e.g. $t^2-u^2$. 
In $t\bar{t}$ plus 1 jet production an asymmetry can already occur at tree
level (essentially, this amounts to a different cut of the same
amplitude). Measurements \cite{Aaltonen:2008hc,Aaltonen:2011kc,Abazov:2007qb,Abazov:2011rq} 
by the Tevatron experiments show
substantial deviations from the Standard Model prediction for pair
production, especially a deviation of more than 3 standard deviations 
by CDF at large invariant $t\bar{t}$ masses \cite{Aaltonen:2011kc}.
For this reason there has been considerable interest in this observable. 

We discuss here the Standard Model calculations for this observable.
A discussion of the many studies of specific New Physics effects on 
the charge or forward-backward asymmetry is beyond the scope of this review.

The effect of this interference can be understood more intuitively
by the statement that the incoming quarks, via the
interference, tend to repel the produced top quarks towards larger
rapidity, and/or attract the produced anti-top quarks toward slightly
smaller rapidities. The net effect, therefore, at the Tevatron, where
the top- anti-top pairs are produced in $q\bar{q}$ annihilation, is a
shift of the top quark rapidity distribution towards larger rapidity,
and of the anti-top distribution towards smaller values. This clearly
creates a $y$-dependent asymmetry of the type (\ref{eq:17}). Because of
these shifts, this also corresponds to a forward-backward asymmetry $A_{\mathrm{FB}}$.

This intuition may also be obtained in threshold resummation
from the so-called soft anomalous dimension in the $q\bar{q}$ channel, which governs
subleading threshold logarithms; leading logarithms are symmetric under
$t \leftrightarrow u$ interchange, and therefore cancel in the
asymmetry. The subleading contribution in the $q\bar{q}$ channel reads \cite{Kidonakis:1996aq}
\begin{equation}
\label{eq:18}
  \Delta \sigma = \exp \left\{\alpha_s L\Big[ \frac{32}{6}-\frac{27}{6}\Big]
\ln \frac{u}{t}\right\} \sigma_{\rm Born}\,,
\end{equation}
where $L$ is the threshold logarithm. This expression, through
$\ln(u/t)$, is indeed anti-symmetric under $t \leftrightarrow u$
interchange.

Since the leading contribution to this effect for NLO pair production involves
a loop diagram, the asymmetry itself is then of leading order
accuracy. The impact of even higher orders is then very
interesting. They have first been estimated from approximate,
resummation based calculations to NLL
\cite{Almeida:2008ug,Kidonakis:2001nj} and NNLL
\cite{Ahrens:2011uf,Kidonakis:2011zn,Kidonakis:2015ona}.
For this only resummations based on
threshold $T_3$, see Eq.~(\ref{eq:20}), can be used, as the other two thresholds
are not sensitive to the top quark rapidity. The higher order
corrections so computed are small,
and reasonably insensitive to scale variations.
Hence, based on these approximate calculations, the discrepancy
would persist, although in recent 
analyses by D0 \cite{Abazov:2013wxa,Chapelain:2013eea} it is found to be not so
large.

For the $t\bar{t}$ case the electroweak corrections have been calculated
\cite{Bernreuther:2012sx,Hollik:2011ps,Kuhn:2011ri}. They are
unexpectedly
large, thus also diminishing the overall discrepancy. 
It is worth noting that, from a slightly different perspective, effects of colour reconnection in
parton shower algorithms can already cause an asymmetry
at what is formally leading order \cite{Skands:2012mm}.

Very recently the exact calculations for the charge asymmetry
to NNLO were completed \cite{Czakon:2014xsa}. Upon taking into account
the second order QCD corrections in addition to the first order EW
corrections a shift of no less than 27\% with respect to the NLO QCD asymmetry was
found, yielding a value of $A_{FB} = 0.095 \pm 0.007$. This is now in 
good agreement with the most recent 
D0 measurement of $0.106 \pm 0.03$ \cite{Abazov:2014cca},
and only somewhat below the CDF \cite{Aaltonen:2012it} value of $0.164\pm 0.047$,
which seems to settle this issue to a large extent.  

Besides defining the asymmetry in terms of the top quark itself
(\ref{eq:17}), one may define it also in terms of the leptons produced
in top and/or anti-top decay , either in the lepton-plus-jets or the di-lepton
channel. The $A^{ll}_{FB}$ asymmetry will be in general a little washed out, but leptons
are relatively easy to measure. There is however still a need
for unfolding due to limited acceptance.
A recent compilation of theory predictions including leptonic
asymmetries is available\cite{Bernreuther:2012sx}. 

At the Tevatron, CDF and D0 have performed a set of measurements for
various types of asymmetries. At the constructed top quark level 
the measured asymmetries exceed the theory prediction by a few
standard deviations. 
We already mentioned the top quark level asymmetries by D0 and CDF. 
Recent $A^{ll}_{FB}$ measurements in the lepton-plus-jets channels corrected to 
the parton level are $16.4\pm 4.7 \%$ (CDF) and 
$12.6\pm 6.5 \%$ (D0), vs. $8.8\pm 0.6\%$ according to the SM.
An overview can be found in Ref.~[\cite{Sharyy:2013dfa}].

As noted above, the charge asymmetry is present at leading order in
$t\bar{t}+\,\mathrm{jet}$ production.  However, here NLO corrections
\cite{Dittmaier:2007wz, Melnikov:2010iu} tend to wash out the asymmetry for
this reaction. An explanation for this effect was given in 
Ref.~[\cite{Melnikov:2010iu}],  based on the following structure of the NLO
forward-backward asymmetry for this reaction
\begin{equation}
  \label{eq:8}
  A_{\mathrm{FB}} (t\bar{t}j) = \alpha_s^3 \frac{C}{\ln(m/p_{T,j})} + \alpha_s^4 D_{\mathrm{hard}}\,.
\end{equation}
The second term, appearing at NLO, cancels the first as they
have opposite signs. The inverse logarithm is due to the fact that the 
denominator in the asymmetry has a higher power of leading soft logarithms.
Also for $t\bar{t}jj$ the NLO term seems to reduce the LO contribution
to the asymmetry \cite{Bevilacqua:2011aa}.

At the LHC, the net effect of the QCD induced asymmetry is an overall broadening of the top quark
rapidity distributions and a slight narrowing of the anti-top rapidity
distribution. Here there is therefore no forward-backward asymmetry,
but a charge asymmetry that is most pronounced at larger rapidities.  One proposal \cite{Hewett:2011wz}
is e.g. to assess the asymmetry using only events with (anti)tops
above a certain minimum rapidity, of about 1.5.  
New observables with promising sensitivity have been proposed\cite{Berge:2012rc,Berge:2013xsa}.

At 7 TeV, a combination of CMS and ATLAS measurements\cite{ATLASchargeasycomb,CMSchargeasycomb}
of the charge asymmetry yields $0.005 \pm 0.007\, (\mathrm{stat}) \pm 0.006\, (\mathrm{syst})$,
in agreement with the NLO QCD and electroweak theory\cite{Kuhn:2011ri},
although also compatible with a lack of asymmetry.


\subsection{Invariant mass and other distributions}
\label{sec:invariant-mass}

Besides inclusive observables such as the cross section and charge
asymmetry, differential distribution afford a more detailed look 
into production dynamics. For instance, a moderate enhancement 
in tails of distributions due to New Physics would possibly not be visible 
in the inclusive observables.  
An important distribution for both the Tevatron and the LHC is
in the invariant mass $M_{t\bar{t}}$. The shape of the distribution in
the SM has a relatively small uncertainty. It has been computed in
approximate NNLO in resummed NNLL accuracy \cite{Ahrens:2010zv}.

It is sensitive to the top mass, and may thus assist in determining
it. Shape deviations from the QCD predictions in this distribution
(peaks, peak-dip structures) are telltales of new physics, such as
resonances with various spin, parity and colour quantum numbers. A
study employing the flexibility of MadGraph in a bottom-up approach
was performed in Ref.~[\cite{Frederix:2007gi}], in which only the most
generic aspects of new models are used.
Given that the exact charge asymmetry calculation \cite{Czakon:2014xsa} was
based on a fully differential NNLO calculation for pair production, various
differential distributions will soon be available at that accuracy. 

Approximate calculations based on resummation methods, discussed earlier, have
already been done, e.g. for the invariant mass distribution
\cite{Ferroglia:2013zwa}, and for single particle inclusive
distributions at the NNLO and beyond level
\cite{Ferroglia:2013awa,Kidonakis:2014pja,Ahrens:2011mw}.

Measurements of differential distributions in variables 
associated with the top quark pair have been performed, as
well as of single particle inclusive distributions
\cite{Aad:2015eia,Khachatryan:2015oqa,Chatrchyan:2012saa,CDF:2013gna,Jung:2012ih}.

\subsection{Single top production}
\label{sec:single-top}

\begin{figure}
  \centerline{\includegraphics[width=0.7\columnwidth]{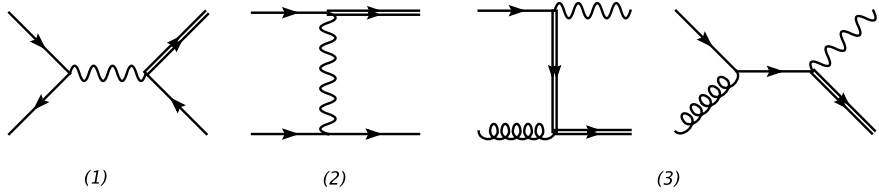}}
  \caption{From left to right the $s$-channel (1), $t$-channel (2) processes,
    and the $Wt$ associated (3) production
    channel.}\label{Fig:singletopprocesses}
\end{figure}

Tops can be produced singly through the weak interaction, in
processes that are customarily categorized 
by names referring to kinematics in the Born approximation, 
see Fig.~\ref{Fig:singletopprocesses}. 
Important aspects of single-top production are 
that $V_{tb}$ can be directly measured without assuming three fermion
generations, and that the chiral
structure of the associated vertex can be tested. The latter is the case
because a single top produced in this way
is highly polarized, which offers a chance to study the chirality of the coupling
via spin correlations, as discussed in eq.~(\ref{eq:22}), 
section \ref{sec:top-quark-decay}.
Another feature is that the dominant $t$-channel at the LHC, when 
confronting measurements with a 5-flavour NLO calculation,
will help determine the $b$-quark density. (In a 4-flavour scheme, one
would demand an extra forward (b) jet). 
Finally, it is interesting that the different single top production processes
are each sensitive to different varieties of new physics.
Thus, the $s$-channel will be sensitive to e.g. $W'$ resonances, the $t$-channel to FCNC's. 
Note that the channel separation according to Fig.~\ref{Fig:singletopprocesses} holds to NLO, 
but not to all orders; at higher orders interference can take place between 
channels. Let us however maintain this separation, and discuss the channels
separately.

\subsubsection{$s$ and $t$ channel}
\label{sec:s-t-channel}

Experimentally, both of these single top
production processes turned out to be rather more difficult to
separate from backgrounds than expected, as the latter are large,
and similar in shape to the signals.
Based on samples of up to 9.7~$\mathrm{fb^{-1}}$ per experiment,
the Tevatron combination\cite{Group:2009qk} of a number of CDF and D0 measurements yields an
inclusive single top production cross section of 
\begin{equation}
\sigma = 3.30^{+0.52}_{-0.40} \,\, \mathrm{pb}\,,
\end{equation}
and a measurement of $|V_{tb}| = 1.02^{+0.06}_{-0.05}$.
Furthermore, CDF and D0 have reported the Tevatron combination\cite{CDF:2014uma} of inclusive single top production
in the $s$ channel only, with a cross section of 
\begin{equation}
\sigma_s = 1.29^{+0.26}_{-0.24} \,\, \mathrm{pb} \,.
\end{equation}
%

At the LHC at a centre-of-mass energy of 7~TeV, 
the inclusive SM production rates of the $s$-channel, $t$-channel and $Wt$ channel
are approximately 4.6, 65 and 16 pb respectively; at 8 TeV they are 5.6, 88 and 22 pb, respectively.
The $t$-channel yields clearly the dominant contribution. Besides interesting
in its own right, the $t$-channel process is a background to many new physics
processes involving both neutral and charged Higgs production.
Based on samples of 4.6 $\mathrm{fb^{-1}}$ by ATLAS and 1.14 $\mathrm{fb^{-1}}$ by CMS
of the run at 7~TeV,
the $t$-channel cross section is~\cite{Aad:2014fwa,Chatrchyan:2012ep}
\begin{align}
\mathrm{ATLAS}\, : & \quad \sigma_t = 68 \pm 2\, \mathrm{(stat)} \pm 8\, \mathrm{(sys)}\, \mathrm{pb}\,,
\nonumber \\
\mathrm{CMS}\, : & \quad \sigma_t = 67.2 \pm 3.7\, \mathrm{(stat)} \pm 4.8\, \mathrm{(sys)}\, \mathrm{pb}\,.
\end{align}
Based on samples of 20.3 $\mathrm{fb^{-1}}$ by ATLAS and 19.7 $\mathrm{fb^{-1}}$ by CMS
of the run at 8~TeV, the $t$-channel cross section is~\cite{ATLAS2014007,Khachatryan:2014iya}
\begin{align}
\mathrm{ATLAS}\, : & \quad \sigma_t = 82.6 \pm 1.2\, \mathrm{(stat)} \pm 12.0\, \mathrm{(sys)}\, \mathrm{pb}\,,
\nonumber \\
\mathrm{CMS}\, : & \quad \sigma_t = 83.6 \pm 2.3\, \mathrm{(stat)} \pm 7.4\, \mathrm{(sys)}\, \mathrm{pb}\,.
\end{align}
A combination, based on partial data samples of 5.8 $\mathrm{fb^{-1}}$ by ATLAS and 5.0 $\mathrm{fb^{-1}}$ by CMS,
yields~\cite{ATLASCMSsigmatcomb}
\begin{equation}
\mathrm{ATLAS/CMS}\quad :
\sigma_t = 85 \pm 4\, \mathrm{(stat)} \pm 11\, \mathrm{(sys)} \pm 3\, \mathrm{(lumi)}\,\, \mathrm{pb}\,,
\end{equation}
with a total uncertainty of 12.1~pb, and in good agreement with the SM prediction.
For all the measurements above, the values of $V_{tb}$ which are extracted are compatible with 1.

Based on the sample of 20.3 $\mathrm{fb^{-1}}$ of the run at 8~TeV, ATLAS has found a first evidence of
$s$-channel production\cite{ATLAS2015047} at 3.2$\sigma$ level
\begin{equation}
\mathrm{ATLAS}\quad :
\sigma_s = 4.8 \pm 1.1\, \mathrm{(stat)} {}^{+ 2.2}_{-2.0}\, \mathrm{(sys)}\,\, \mathrm{pb}\,.
\end{equation}

On the theory side, the single top cross section has been computed at NLO accuracy in the QCD and electroweak 
corrections~\cite{Bordes:1994ki,Pittau:1996rp,Stelzer:1997ns,Harris:2002md,Cao:2004ap,Cao:2005pq,Campbell:2004ch,
Campbell:2005bb,Beccaria:2008av,Campbell:2009gj,Heim:2009ku,Falgari:2010sf,Schwienhorst:2010je,
Falgari:2011qa,Papanastasiou:2013dta}, including 
resummations~\cite{Kidonakis:2006bu,Wang:2010ue,Zhu:2010mr,Kidonakis:2011wy} and
matching NLO computations to parton showers~\cite{Frixione:2005vw,Alioli:2009je,Frederix:2012dh};
and at NNLO accuracy in the QCD corrections~\cite{Brucherseifer:2014ama}.
The NLO (and NNLO) corrections are at a few percent level, and within errors
the measured cross sections agree with them. Approximate NNLO $p_T$
distributions have recently appeared \cite{Wang:2012dc,Kidonakis:2015wva}.

\subsubsection{Wt associated production}
\label{sec:wt}

\begin{figure}[t]
\centering
\includegraphics[width=0.55\columnwidth]{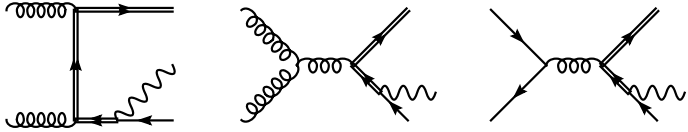}
\caption{Doubly resonant diagrams in NLO corrections to $Wt$ production.}\label{fig:wtlo} 
\end{figure}

A subtle and interesting issue arises in the $Wt$ mode of single top production at NLO. 
In the radiative corrections some diagrams contain an
intermediate anti-top decaying into a $W$ and anti-down type quark, that can become resonant. 
From another viewpoint, these diagrams can be interpreted as LO $t\bar{t}$ on-shell
production, with subsequent $\bar{t}$ decay, see Fig.~\ref{fig:wtlo}.
One is therefore faced with the issue to what extent the $Wt$ and
$t\bar{t}$ can be properly defined and separated as individual processes. 
To this end several definitions of the $Wt$ channel have been given in the literature,
each with the aim of recovering a well-behaved expansion in $\alpha_s$. This problem
of interference is of course not uncommon in computations
at order of at least ${\cal O}(g_w^2\alpha_s^2)$. The
cross section at this order has been previously presented in
Refs.~[\cite{Tait:1999cf,Belyaev:1998dn,Kersevan:2006fq}], where only
tree-level graphs were considered, and in 
Refs.~[\cite{Zhu:2002uj,Campbell:2005bb,Cao:2008af}], where one-loop
contributions were included as well. The rather vexing point here is
that the $t\bar{t}$ process with which the $Wt$ interferes is an order of
magnitude larger, rendering the NLO correction much too large.

In Ref.~[\cite{Frixione:2008yi}]  this interference issue was addressed
extensively in the context of event generation, in particular the
\texttt{MC@NLO} framework (\texttt{POWHEG} has implemented 
the same method \cite{Re:2010bp}). Two different procedures for subtracting the doubly-resonant contributions
and thereby recovering a perturbatively well-behaved $Wt$ cross section
were defined. In ``Diagram Removal (DR)''  the graphs in 
Fig.~\ref{fig:wtlo} were eliminated from the calculation, while in ``Diagram Subtraction (DS)''
the doubly resonant contribution was removed via a judiciously constructed subtraction term.
The DS procedure leads to the following expression for the cross section
\begin{equation}
  \label{eq:5}
  d\sigma^{(2)} + \sum_{\alpha\beta} \int \frac{dx_1dx_2}{x_1x_2S}
\mathcal{L}_{\alpha\beta} \left(\hat{S}_{\alpha\beta} + I_{\alpha\beta}
+ D_{\alpha\beta}-\tilde{D}_{\alpha\beta}  \right) d\phi_3,
\end{equation}
where $\alpha\beta$ labels the initial state channel in which
the doubly-resonant contribution occurs: $gg \; \mathrm{or} \; q\bar{q}$.
$\hat{S}$ is the square of the non-resonant diagrams, 
$I$ their interference with $D$, the square of 
graphs of Fig.~\ref{fig:wtlo}. The subtraction term $\tilde{D}$ requires
careful construction \cite{Frixione:2008yi}.
It was shown that, with suitable cuts, the interference terms are small.
From Eq.~(\ref{eq:5}) one sees that the
difference of DR and DS in essence consists of the interference term.
A particularly suitable cut is imposing a maximum on
the $p_T$ of the second hardest $b$-flavoured hadron, 
a generalization of a proposal made in Ref.~[\cite{Campbell:2005bb}].
Thus defined, the $Wt$ and $t\bar{t}$ cross sections
can be separatedly considered to NLO. 

The experimental status of this production mode at present
is as follows.  In the 7~TeV run, ATLAS \cite{Aad:2012xca} and CMS
\cite{Chatrchyan:2012zca} have measured the $Wt$-channel cross
section, with the results
\begin{align}
\mathrm{ATLAS} [2.05\,\mathrm{fb^{-1}}]: & \quad \sigma_{Wt} = 16.8 \pm 2.9\, \mathrm{(stat)} \pm 4.9 \mathrm{(sys)}\, \mathrm{pb}\,,
\nonumber \\
\mathrm{CMS}\, [4.9\,\mathrm{fb^{-1}}]: & \quad \sigma_{Wt} = 16^{+5}_{-4}\, \mathrm{pb}\,.
\end{align}
In the 8~TeV run, a combination of $Wt$-channel measurements has been performed\cite{Aad:2015eto,CMSPASTOP14009},
based on a data set of 12.2 $\mathrm{fb^{-1}}$ by CMS and 20.3 $\mathrm{fb^{-1}}$ by ATLAS
\begin{equation}
\mathrm{ATLAS/CMS}: \quad \sigma_{Wt} = 25.0 \pm 1.4\, \mathrm{(stat)} \pm 4.4 \mathrm{(sys)} \pm 0.7 \mathrm{(lumi)}\, \mathrm{pb}\,,
\end{equation} 
with a total uncertainty of 4.7~pb.

Within errors, the $Wt$-channel cross section measurements above agree with the NLO 
calculations\cite{Zhu:2002uj,Campbell:2005bb,Cao:2008af}, approximate
NNLO \cite{Kidonakis:2010ux} and
the NLO plus parton showers discussed above \cite{Frixione:2008yi,White:2009yt,Re:2010bp}.

One way to avoid the above difficulties in separating $Wt$ from
$\bar{t}t$ is to consider the common final state $WWbb$ (in the
4-flavour scheme) and not ask
if there were one or two intermediate top quarks involved in producing
this final state -- zero intermediate top quarks is also a possibility here.
In Refs.~[\cite{Frederix:2013gra,Cascioli:2013wga}] a unified approach 
in the 4-flavour scheme was taken in which both $Wt$ and 
$\bar{t}t$ produce as final state $WWbb$, and the NLO corrections were
computed.

\subsection{Associated top production at higher order}
\label{sec:assoc-prod}

One can also consider processes where a top pair 
or a single top are produced in association with other particles.
Given its relevance for the measurement of the Higgs-top Yukawa coupling,
clearly the most important associated production is the production of a top pair in
association with a Higgs boson, $t\bar{t} h$. This process is known to NLO accuracy
in QCD, at parton level~\cite{Beenakker:2002nc,Dawson:2003zu} and interfaced to
parton showers~\cite{Frederix:2011zi,Garzelli:2011vp}. Also the electroweak
corrections have been computed~\cite{Frixione:2014qaa,Frixione:2015zaa}.

The rapid evolution of computations of scattering processes with many
final-state particles to NLO accuracy in QCD -- the so-called NLO revolution --
has left its mark on processes involving top production as
well, yielding calculations that would have been hard to imagine some years ago.
Accordingly, many important backgrounds to top pair production in association with a Higgs boson,
with subsequent decay of the Higgs boson into a bottom-quark pair or into a pair of photons,
have been computed to NLO accuracy. In particular,
production of a top pair in association with a jet is known to NLO accuracy at parton level~\cite{Dittmaier:2007wz,Melnikov:2010iu}, 
including the top decays~\cite{Melnikov:2011qx}, as well as non-resonant diagrams, interferences and off-shell 
effects\cite{Bevilacqua:2015qha}, and interfaced with parton showers~\cite{Kardos:2011qa,Alioli:2011as,Czakon:2015cla};
top pair production in association with two jets is known to NLO accuracy at parton level~\cite{Bevilacqua:2010ve,Bevilacqua:2011aa},
and interfaced with parton showers~\cite{Hoeche:2014qda}; production of a top pair in association with a $b\bar{b}$ pair
is known at parton level~\cite{Bredenstein:2009aj,Bevilacqua:2009zn,Bredenstein:2010rs} and interfaced with parton 
showers~\cite{Kardos:2013vxa,Cascioli:2013era,Garzelli:2014aba};
production of a top pair in association with a photon is known to NLO accuracy at parton level~\cite{Melnikov:2011ta}
and interfaced with parton showers~\cite{Alwall:2014hca,Kardos:2014zba};
production of a top pair in association with two photons is known to NLO accuracy at parton level and interfaced with parton 
showers~\cite{Alwall:2014hca,Kardos:2014pba}.

Furthermore, the production of a top pair in association with a $Z$~boson is known to NLO accuracy 
at parton level~\cite{Lazopoulos:2008de,Kardos:2011na}
and interfaced with parton
showers~\cite{Garzelli:2011is,Garzelli:2012bn}, which is relevant to
measure the $t\bar{t} Z$ coupling \cite{Rontsch:2015una,Rontsch:2014cca};
the production of a top pair in association with a $W$~boson is known to NLO accuracy~\cite{Badger:2010mg,Campbell:2012dh,Maltoni:2014zpa} 
and interfaced with parton showers~\cite{Garzelli:2012bn}, which can be used as a tool to examine the top-quark charge asymmetry.
Also the electroweak corrections to the production of a top pair in association with a $W/Z$~boson have been computed~\cite{Frixione:2015zaa}.
The production of a top pair in association with two vector bosons, be it either $W,\, Z$ bosons or photons,
is known to NLO accuracy at parton level, and interfaced with parton showers~\cite{Alwall:2014hca,Maltoni:2015ena};
as well as the production of a top pair in association with a vector boson and a jet 
is known to NLO accuracy at parton level, and interfaced with parton showers~\cite{Alwall:2014hca}.
Finally, the production of two top pairs is known to NLO accuracy at parton level~\cite{Bevilacqua:2012em,Maltoni:2015ena},
which can be used as a benchmark process to test New Physics signals.

In addition, single top production in association with a $W$ boson is known to 
NLO accuracy at parton level~\cite{Frixione:2008yi,Cascioli:2013wga},
and interfaced to parton showers~\cite{Frixione:2008yi},
which is relevant for the $W t$ mode of single top production, see Sect.~\ref{sec:wt};
single top production in association with a $Z$ boson is known to 
NLO accuracy at parton level~\cite{Campbell:2013yla},
which is a background to flavour changing neutral current decays of the top in $t{\bar t}$ production.
Various single top production processes in association with a $b$ quark and a $Z$ boson or a photon
or a jet are available to NLO accuracy in Ref.~[\cite{Alwall:2014hca}].

\section{Conclusions}
\label{sec:conclusions}

Top quark physics is at present at a pivotal point, 
in the early days of Run II of the LHC. Rather accurate
studies of top quark observables from Tevatron and LHC Run I 
data have been done, but the bulk of (higher energy) data 
is still to be collected. Also in top physics the Standard
Model has withstood tests so far, but many highly detailed and varied 
tests by the LHC experiments will follow. 

Top's attractiveness as a study object has by 
no means diminished. On the contrary, new observables
are being enlisted for this enterprise. The characteristics
of production and decay, in association with other
particles, can be very revealing. The examination of 
many (multi-)differential distributions,
a full accounting of spin and off-shellness, and especially
its interaction with the Higgs boson are all still to come.

As we have reviewed here, the theoretical tools for top physics studies 
are of high quality, and still keep improving with remarkable pace. We are
therefore confident that the top quark will remain in the focus
of attention for a good many more years.

\section*{Acknowledgments}

We would like to thank F. Maltoni and M. Vreeswijk for valuable
discussions and comments.
EL has been supported by the Netherlands Foundation for Fundamental
Research of Matter (FOM) programme 156, entitled  ``Higgs as Probe and Portal'',
and the National Organization for Scientific Research (NWO). 
This work was also supported by the Research Executive Agency (REA)
of the European Union under Grant Agreement numbers
PITN-GA2012-316704 (HiggsTools), and
PITN-GA-2010-264564 (LHCPhenoNet).

\bibliographystyle{utphys}
\bibliography{references.bib}

\providecommand{\href}[2]{#2}\begingroup\raggedright\begin{thebibliography}{100}

\bibitem{Abe:1995hr}
{\bf CDF} Collaboration, F.~Abe {\em et.~al.}, ``Observation of top quark
  production in anti-p p collisions,'' {\em Phys. Rev. Lett.} {\bf 74} (1995)
  2626--2631,
\href{http://xxx.lanl.gov/abs/hep-ex/9503002}{{\tt hep-ex/9503002}}.

\bibitem{Abachi:1995iq}
{\bf D0} Collaboration, S.~Abachi {\em et.~al.}, ``Observation of the top
  quark,'' {\em Phys. Rev. Lett.} {\bf 74} (1995) 2632--2637,
\href{http://xxx.lanl.gov/abs/hep-ex/9503003}{{\tt hep-ex/9503003}}.

\bibitem{Glashow:1970gm}
S.~L. Glashow, J.~Iliopoulos, and L.~Maiani, ``{Weak Interactions with
  Lepton-Hadron Symmetry},'' {\em Phys. Rev.} {\bf D2} (1970)
1285--1292.

\bibitem{Kobayashi:1973fv}
M.~Kobayashi and T.~Maskawa, ``{CP Violation in the Renormalizable Theory of
  Weak Interaction},'' {\em Prog. Theor. Phys.} {\bf 49} (1973)
652--657.

\bibitem{Bernreuther:2008ju}
W.~Bernreuther, ``{Top quark physics at the LHC},'' {\em J.Phys.} {\bf G35}
  (2008) 083001,
\href{http://xxx.lanl.gov/abs/0805.1333}{{\tt 0805.1333}}.

\bibitem{Han:2008xb}
T.~Han, ``{The 'Top Priority' at the LHC},'' {\em Int.J.Mod.Phys.} {\bf A23}
  (2008) 4107--4124,
\href{http://xxx.lanl.gov/abs/0804.3178}{{\tt 0804.3178}}.

\bibitem{Quadt:2007jk}
A.~Quadt, ``{Top quark physics at hadron colliders},'' {\em Eur.Phys.J.} {\bf
  C48} (2006)
835--1000.

\bibitem{Schilling:2012dx}
F.-P. Schilling, ``{Top Quark Physics at the LHC: A Review of the First Two
  Years},'' {\em Int. J. Mod. Phys.} {\bf A27} (2012) 1230016,
\href{http://xxx.lanl.gov/abs/1206.4484}{{\tt 1206.4484}}.

\bibitem{JABEEN:2013mva}
S.~Jabeen, ``{Top and Higgs Physics at the Hadron Colliders},'' {\em Int. J.
  Mod. Phys.} {\bf A28} (2013)
1330038.

\bibitem{'tHooft:1979bh}
G.~'t~Hooft, ``{Naturalness, chiral symmetry, and spontaneous chiral symmetry
  breaking},'' {\em NATO Adv.Study Inst.Ser.B Phys.} {\bf 59} (1980)
135.

\bibitem{Ibanez:1982fr}
L.~E. Ibanez and G.~G. Ross, ``{SU(2)-L x U(1) Symmetry Breaking as a Radiative
  Effect of Supersymmetry Breaking in Guts},'' {\em Phys. Lett.} {\bf B110}
  (1982)
215--220.

\bibitem{Branchina:2013jra}
V.~Branchina and E.~Messina, ``{Stability, Higgs Boson Mass and New Physics},''
  {\em Phys. Rev. Lett.} {\bf 111} (2013) 241801,
\href{http://xxx.lanl.gov/abs/1307.5193}{{\tt 1307.5193}}.

\bibitem{Bezrukov:2012sa}
F.~Bezrukov, M.~{\relax Yu}. Kalmykov, B.~A. Kniehl, and M.~Shaposhnikov,
  ``{Higgs Boson Mass and New Physics},'' {\em JHEP} {\bf 10} (2012) 140,
\href{http://xxx.lanl.gov/abs/1205.2893}{{\tt 1205.2893}}.

\bibitem{Degrassi:2012ry}
G.~Degrassi, S.~Di~Vita, J.~Elias-Miro, J.~R. Espinosa, G.~F. Giudice,
  G.~Isidori, and A.~Strumia, ``{Higgs mass and vacuum stability in the
  Standard Model at NNLO},'' {\em JHEP} {\bf 08} (2012) 098,
\href{http://xxx.lanl.gov/abs/1205.6497}{{\tt 1205.6497}}.

\bibitem{Buttazzo:2013uya}
D.~Buttazzo, G.~Degrassi, P.~P. Giardino, G.~F. Giudice, F.~Sala, A.~Salvio,
  and A.~Strumia, ``{Investigating the near-criticality of the Higgs boson},''
  {\em JHEP} {\bf 12} (2013) 089,
\href{http://xxx.lanl.gov/abs/1307.3536}{{\tt 1307.3536}}.

\bibitem{Alekhin:2012py}
S.~Alekhin, A.~Djouadi, and S.~Moch, ``{The top quark and Higgs boson masses
  and the stability of the electroweak vacuum},'' {\em Phys. Lett.} {\bf B716}
  (2012) 214--219,
\href{http://xxx.lanl.gov/abs/1207.0980}{{\tt 1207.0980}}.

\bibitem{Abazov:2012vd}
{\bf D0} Collaboration, V.~M. Abazov {\em et.~al.}, ``{An Improved
  determination of the width of the top quark},'' {\em Phys.Rev.} {\bf D85}
  (2012) 091104,
\href{http://xxx.lanl.gov/abs/1201.4156}{{\tt 1201.4156}}.

\bibitem{Aaltonen:2013kna}
{\bf CDF} Collaboration, T.~A. Aaltonen {\em et.~al.}, ``{Direct Measurement of
  the Total Decay Width of the Top Quark},'' {\em Phys. Rev. Lett.} {\bf 111}
  (2013), no.~20, 202001,
\href{http://xxx.lanl.gov/abs/1308.4050}{{\tt 1308.4050}}.

\bibitem{Frixione:2007zp}
S.~Frixione, E.~Laenen, P.~Motylinski, and B.~R. Webber, ``{Angular
  correlations of lepton pairs from vector boson and top quark decays in Monte
  Carlo simulations},'' {\em JHEP} {\bf 0704} (2007) 081,
\href{http://xxx.lanl.gov/abs/hep-ph/0702198}{{\tt hep-ph/0702198}}.

\bibitem{Artoisenet:2012st}
P.~Artoisenet, R.~Frederix, O.~Mattelaer, and R.~Rietkerk, ``{Automatic
  spin-entangled decays of heavy resonances in Monte Carlo simulations},'' {\em
  JHEP} {\bf 03} (2013) 015,
\href{http://xxx.lanl.gov/abs/1212.3460}{{\tt 1212.3460}}.

\bibitem{Motylinski:2009kt}
P.~Motylinski, ``{Angular correlations in t-channel single top production at
  the LHC},'' {\em Phys. Rev.} {\bf D80} (2009) 074015,
\href{http://xxx.lanl.gov/abs/0905.4754}{{\tt 0905.4754}}.

\bibitem{Melnikov:2009dn}
K.~Melnikov and M.~Schulze, ``{NLO QCD corrections to top quark pair production
  and decay at hadron colliders},'' {\em JHEP} {\bf 08} (2009) 049,
\href{http://xxx.lanl.gov/abs/0907.3090}{{\tt 0907.3090}}.

\bibitem{Denner:2012yc}
A.~Denner, S.~Dittmaier, S.~Kallweit, and S.~Pozzorini, ``{NLO QCD corrections
  to off-shell top-antitop production with leptonic decays at hadron
  colliders},'' {\em JHEP} {\bf 10} (2012) 110,
\href{http://xxx.lanl.gov/abs/1207.5018}{{\tt 1207.5018}}.

\bibitem{Papanastasiou:2013dta}
A.~Papanastasiou, R.~Frederix, S.~Frixione, V.~Hirschi, and F.~Maltoni,
  ``{Single-top $t$-channel production with off-shell and non-resonant
  effects},'' {\em Phys.Lett.} {\bf B726} (2013) 223--227,
\href{http://xxx.lanl.gov/abs/1305.7088}{{\tt 1305.7088}}.

\bibitem{Falgari:2011qa}
P.~Falgari, F.~Giannuzzi, P.~Mellor, and A.~Signer, ``{Off-shell effects for
  t-channel and s-channel single-top production at NLO in QCD},'' {\em
  Phys.Rev.} {\bf D83} (2011) 094013,
\href{http://xxx.lanl.gov/abs/1102.5267}{{\tt 1102.5267}}.

\bibitem{Alwall:2014hca}
J.~Alwall, R.~Frederix, S.~Frixione, V.~Hirschi, F.~Maltoni, {\em et.~al.},
  ``{The automated computation of tree-level and next-to-leading order
  differential cross sections, and their matching to parton shower
  simulations},'' {\em JHEP} {\bf 1407} (2014) 079,
\href{http://xxx.lanl.gov/abs/1405.0301}{{\tt 1405.0301}}.

\bibitem{Denner:2006ic}
A.~Denner and S.~Dittmaier, ``{The Complex-mass scheme for perturbative
  calculations with unstable particles},'' {\em Nucl. Phys. Proc. Suppl.} {\bf
  160} (2006) 22--26, \href{http://xxx.lanl.gov/abs/hep-ph/0605312}{{\tt
  hep-ph/0605312}}.
[,22(2006)].

\bibitem{Denner:2005fg}
A.~Denner, S.~Dittmaier, M.~Roth, and L.~H. Wieders, ``{Electroweak corrections
  to charged-current e+ e- ---> 4 fermion processes: Technical details and
  further results},'' {\em Nucl. Phys.} {\bf B724} (2005) 247--294,
  \href{http://xxx.lanl.gov/abs/hep-ph/0505042}{{\tt hep-ph/0505042}}.
[Erratum: Nucl. Phys.B854,504(2012)].

\bibitem{Denner:1999gp}
A.~Denner, S.~Dittmaier, M.~Roth, and D.~Wackeroth, ``{Predictions for all
  processes e+ e- ---> 4 fermions + gamma},'' {\em Nucl. Phys.} {\bf B560}
  (1999) 33--65,
\href{http://xxx.lanl.gov/abs/hep-ph/9904472}{{\tt hep-ph/9904472}}.

\bibitem{Heinrich:2013qaa}
G.~Heinrich, A.~Maier, R.~Nisius, J.~Schlenk, and J.~Winter, ``{NLO QCD
  corrections to $W^{+} W^{-}b\bar{b}$ production with leptonic decays in the
  light of top quark mass and asymmetry measurements},'' {\em JHEP} {\bf 06}
  (2014) 158,
\href{http://xxx.lanl.gov/abs/1312.6659}{{\tt 1312.6659}}.

\bibitem{Kane:1991bg}
G.~L. Kane, G.~A. Ladinsky, and C.~P. Yuan, ``{Using the Top Quark for Testing
  Standard Model Polarization and CP Predictions},'' {\em Phys. Rev.} {\bf D45}
  (1992)
124--141.

\bibitem{Chanowitz:1985hj}
M.~S. Chanowitz and M.~K. Gaillard, ``{The TeV Physics of Strongly Interacting
  W's and Z's},'' {\em Nucl. Phys.} {\bf B261} (1985)
379.

\bibitem{Peskin:1995ev}
M.~E. Peskin and D.~V. Schroeder, {\em {An Introduction to quantum field
  theory}}.
\newblock
1995.
\newblock

\bibitem{Czarnecki:2010gb}
A.~Czarnecki, J.~G. Korner, and J.~H. Piclum, ``{Helicity fractions of W bosons
  from top quark decays at NNLO in QCD},'' {\em Phys. Rev.} {\bf D81} (2010)
  111503,
\href{http://xxx.lanl.gov/abs/1005.2625}{{\tt 1005.2625}}.

\bibitem{Aaltonen:2012rz}
{\bf CDF, D0} Collaboration, T.~Aaltonen {\em et.~al.}, ``{Combination of CDF
  and D0 measurements of the $W$ boson helicity in top quark decays},'' {\em
  Phys. Rev.} {\bf D85} (2012) 071106,
\href{http://xxx.lanl.gov/abs/1202.5272}{{\tt 1202.5272}}.

\bibitem{Aaltonen:2012lua}
{\bf CDF} Collaboration, T.~Aaltonen {\em et.~al.}, ``{Measurement of $W$-Boson
  Polarization in Top-quark Decay using the Full CDF Run II Data Set},'' {\em
  Phys. Rev.} {\bf D87} (2013), no.~3, 031104,
\href{http://xxx.lanl.gov/abs/1211.4523}{{\tt 1211.4523}}.

\bibitem{Chatrchyan:2013jna}
{\bf CMS} Collaboration, S.~Chatrchyan {\em et.~al.}, ``{Measurement of the
  W-boson helicity in top-quark decays from $t\bar{t}$ production in
  lepton+jets events in pp collisions at $\sqrt{s} =$ 7 TeV},'' {\em JHEP} {\bf
  10} (2013) 167,
\href{http://xxx.lanl.gov/abs/1308.3879}{{\tt 1308.3879}}.

\bibitem{Khachatryan:2014vma}
{\bf CMS} Collaboration, V.~Khachatryan {\em et.~al.}, ``{Measurement of the W
  boson helicity in events with a single reconstructed top quark in pp
  collisions at $ \sqrt{s}=8 $ TeV},'' {\em JHEP} {\bf 01} (2015) 053,
\href{http://xxx.lanl.gov/abs/1410.1154}{{\tt 1410.1154}}.

\bibitem{Tevatron:2014cka}
{\bf CDF Collaboration, D0} Collaboration, T.~E.~W. Group, ``{Combination of
  CDF and D0 results on the mass of the top quark using up to 9.7 fb$^{-1}$ at
  the Tevatron},''
\href{http://xxx.lanl.gov/abs/1407.2682}{{\tt 1407.2682}}.

\bibitem{ATLASCMStopmasscomb}
{\bf CMS and ATLAS} Collaboration, CMS-PAS-TOP-13-005 and ATLAS-CONF-2013-102,
  ``{Combination of ATLAS and CMS results on the mass of the top quark using up
  to 4.9 $fb^{-1}$ of data},''.

\bibitem{Khachatryan:2015hba}
{\bf CMS} Collaboration, V.~Khachatryan {\em et.~al.}, ``{Measurement of the
  top quark mass using proton-proton data at $\sqrt{s}$ = 7 and 8 TeV},''
\href{http://xxx.lanl.gov/abs/1509.04044}{{\tt 1509.04044}}.

\bibitem{Baak:2012kk}
M.~Baak, M.~Goebel, J.~Haller, A.~Hoecker, D.~Kennedy, {\em et.~al.}, ``{The
  Electroweak Fit of the Standard Model after the Discovery of a New Boson at
  the LHC},'' {\em Eur.Phys.J.} {\bf C72} (2012) 2205,
\href{http://xxx.lanl.gov/abs/1209.2716}{{\tt 1209.2716}}.

\bibitem{Beneke:1994sw}
M.~Beneke and V.~M. Braun, ``Heavy quark effective theory beyond perturbation
  theory: Renormalons, the pole mass and the residual mass term,'' {\em Nucl.
  Phys.} {\bf B426} (1994) 301--343,
\href{http://xxx.lanl.gov/abs/hep-ph/9402364}{{\tt hep-ph/9402364}}.

\bibitem{Bigi:1994em}
I.~I. Bigi, M.~A. Shifman, N.~G. Uraltsev, and A.~I. Vainshtein, ``The pole
  mass of the heavy quark. perturbation theory and beyond,'' {\em Phys. Rev.}
  {\bf D50} (1994) 2234--2246,
\href{http://xxx.lanl.gov/abs/hep-ph/9402360}{{\tt hep-ph/9402360}}.

\bibitem{Smith:1997xz}
M.~C. Smith and S.~S. Willenbrock, ``Top quark pole mass,'' {\em Phys. Rev.
  Lett.} {\bf 79} (1997) 3825--3828,
\href{http://xxx.lanl.gov/abs/hep-ph/9612329}{{\tt hep-ph/9612329}}.

\bibitem{Chatrchyan:2012ea}
{\bf CMS} Collaboration, S.~Chatrchyan {\em et.~al.}, ``{Measurement of the
  top-quark mass in $t\bar{t}$ events with dilepton final states in $pp$
  collisions at $\sqrt{s}=7$ TeV},'' {\em Eur.Phys.J.} {\bf C72} (2012) 2202,
\href{http://xxx.lanl.gov/abs/1209.2393}{{\tt 1209.2393}}.

\bibitem{ATLAStopmassdilepton}
{\bf ATLAS} Collaboration, ATLAS-CONF-2013-077, ``{Measurement of the Top Quark
  Mass in Dileptonic Top Quark Pair Decays with $\sqrt{s}=7$ TeV ATLAS
  Data},''.

\bibitem{ATLAS:2012aj}
{\bf ATLAS} Collaboration, G.~Aad {\em et.~al.}, ``{Measurement of the top
  quark mass with the template method in the $t \bar{t}\rightarrow$ lepton +
  jets channel using ATLAS data},'' {\em Eur.Phys.J.} {\bf C72} (2012) 2046,
\href{http://xxx.lanl.gov/abs/1203.5755}{{\tt 1203.5755}}.

\bibitem{Chatrchyan:2012cz}
{\bf CMS} Collaboration, S.~Chatrchyan {\em et.~al.}, ``{Measurement of the
  top-quark mass in $t\bar{t}$ events with lepton+jets final states in $pp$
  collisions at $\sqrt{s}=7$ TeV},'' {\em JHEP} {\bf 1212} (2012) 105,
\href{http://xxx.lanl.gov/abs/1209.2319}{{\tt 1209.2319}}.

\bibitem{Chatrchyan:2013xza}
{\bf CMS} Collaboration, S.~Chatrchyan {\em et.~al.}, ``{Measurement of the
  top-quark mass in all-jets $t\bar{t}$ events in pp collisions at $\sqrt{s}$=7
  TeV},''
\href{http://xxx.lanl.gov/abs/1307.4617}{{\tt 1307.4617}}.

\bibitem{ATLAStopmassalllepton}
{\bf ATLAS} Collaboration, ATLAS-CONF-2012-030, ``{Determination of the Top
  Quark Mass with a Template Method in the All-Hadronic Decay Channel using
  2.04 $fb^{-1}$ of ATLAS Data},''.

\bibitem{Hoang:2008xm}
A.~H. Hoang and I.~W. Stewart, ``{Top Mass Measurements from Jets and the
  Tevatron Top-Quark Mass},'' {\em Nucl.Phys.Proc.Suppl.} {\bf 185} (2008)
  220--226, \href{http://xxx.lanl.gov/abs/0808.0222}{{\tt 0808.0222}}.

\bibitem{Melnikov:2000qh}
K.~Melnikov and T.~v. Ritbergen, ``{The Three loop relation between the MS-bar
  and the pole quark masses},'' {\em Phys.Lett.} {\bf B482} (2000) 99--108,
\href{http://xxx.lanl.gov/abs/hep-ph/9912391}{{\tt hep-ph/9912391}}.

\bibitem{Marquard:2015qpa}
P.~Marquard, A.~V. Smirnov, V.~A. Smirnov, and M.~Steinhauser, ``{Quark Mass
  Relations to Four-Loop Order in Perturbative QCD},'' {\em Phys. Rev. Lett.}
  {\bf 114} (2015), no.~14, 142002,
\href{http://xxx.lanl.gov/abs/1502.01030}{{\tt 1502.01030}}.

\bibitem{Langenfeld:2009wd}
U.~Langenfeld, S.~Moch, and P.~Uwer, ``{Measuring the running top-quark
  mass},'' {\em Phys. Rev.} {\bf D80} (2009) 054009,
\href{http://xxx.lanl.gov/abs/0906.5273}{{\tt 0906.5273}}.

\bibitem{Abazov:2011pta}
{\bf D0} Collaboration, V.~M. Abazov {\em et.~al.}, ``{Determination of the
  pole and MSbar masses of the top quark from the $t\bar{t}$ cross section},''
  {\em Phys.Lett.} {\bf B703} (2011) 422--427,
\href{http://xxx.lanl.gov/abs/1104.2887}{{\tt 1104.2887}}.

\bibitem{Ahrens:2010zv}
V.~Ahrens, A.~Ferroglia, M.~Neubert, B.~D. Pecjak, and L.~L. Yang,
  ``{Renormalization-Group Improved Predictions for Top-Quark Pair Production
  at Hadron Colliders},'' {\em JHEP} {\bf 09} (2010) 097,
\href{http://xxx.lanl.gov/abs/1003.5827}{{\tt 1003.5827}}.

\bibitem{Moch:2008qy}
S.~Moch and P.~Uwer, ``{Theoretical status and prospects for top-quark pair
  production at hadron colliders},'' {\em Phys. Rev.} {\bf D78} (2008) 034003,
\href{http://xxx.lanl.gov/abs/0804.1476}{{\tt 0804.1476}}.

\bibitem{Beneke:1999qg}
M.~Beneke, A.~Signer, and V.~A. Smirnov, ``{Top quark production near threshold
  and the top quark mass},'' {\em Phys.Lett.} {\bf B454} (1999) 137--146,
\href{http://xxx.lanl.gov/abs/hep-ph/9903260}{{\tt hep-ph/9903260}}.

\bibitem{Hoang:1999zc}
A.~Hoang and T.~Teubner, ``{Top quark pair production close to threshold: Top
  mass, width and momentum distribution},'' {\em Phys.Rev.} {\bf D60} (1999)
  114027,
\href{http://xxx.lanl.gov/abs/hep-ph/9904468}{{\tt hep-ph/9904468}}.

\bibitem{Hoang:2000yr}
A.~Hoang, M.~Beneke, K.~Melnikov, T.~Nagano, A.~Ota, {\em et.~al.}, ``{Top -
  anti-top pair production close to threshold: Synopsis of recent NNLO
  results},'' {\em Eur.Phys.J.direct} {\bf C2} (2000) 1,
\href{http://xxx.lanl.gov/abs/hep-ph/0001286}{{\tt hep-ph/0001286}}.

\bibitem{Juste:2013dsa}
A.~Juste, S.~Mantry, A.~Mitov, A.~Penin, P.~Skands, {\em et.~al.},
  ``{Determination of the top quark mass circa 2013: methods, subtleties,
  perspectives},''
\href{http://xxx.lanl.gov/abs/1310.0799}{{\tt 1310.0799}}.

\bibitem{Beneke:1998rk}
M.~Beneke, ``A quark mass definition adequate for threshold problems,'' {\em
  Phys. Lett.} {\bf B434} (1998) 115--125,
\href{http://xxx.lanl.gov/abs/hep-ph/9804241}{{\tt hep-ph/9804241}}.

\bibitem{Kondo:1988yd}
K.~Kondo, ``{Dynamical likelihood method for reconstruction of events with
  missing momentum. 1: method and toy models},'' {\em J.Phys.Soc.Jap.} {\bf 57}
  (1988)
4126--4140.

\bibitem{Kondo:1991dw}
K.~Kondo, ``{Dynamical likelihood method for reconstruction of events with
  missing momentum. 2: Mass spectra for $2 \rightarrow 2$ processes},'' {\em
  J.Phys.Soc.Jap.} {\bf 60} (1991)
836--844.

\bibitem{Dalitz:1991wa}
R.~Dalitz and G.~R. Goldstein, ``{The Decay and polarization properties of the
  top quark},'' {\em Phys.Rev.} {\bf D45} (1992)
1531--1543.

\bibitem{Almeida:2010pa}
L.~G. Almeida, S.~J. Lee, G.~Perez, G.~Sterman, and I.~Sung, ``{Template
  Overlap Method for Massive Jets},'' {\em Phys.Rev.} {\bf D82} (2010) 054034,
\href{http://xxx.lanl.gov/abs/1006.2035}{{\tt 1006.2035}}.

\bibitem{Alwall:2010cq}
J.~Alwall, A.~Freitas, and O.~Mattelaer, ``{The Matrix Element Method and QCD
  Radiation},'' {\em Phys.Rev.} {\bf D83} (2011) 074010,
\href{http://xxx.lanl.gov/abs/1010.2263}{{\tt 1010.2263}}.

\bibitem{Campbell:2012cz}
J.~M. Campbell, W.~T. Giele, and C.~Williams, ``{The Matrix Element Method at
  Next-to-Leading Order},'' {\em JHEP} {\bf 1211} (2012) 043,
\href{http://xxx.lanl.gov/abs/1204.4424}{{\tt 1204.4424}}.

\bibitem{Kharchilava:1999yj}
A.~Kharchilava, ``{Top mass determination in leptonic final states with
  $J/\psi$},'' {\em Phys.Lett.} {\bf B476} (2000) 73--78,
\href{http://xxx.lanl.gov/abs/hep-ph/9912320}{{\tt hep-ph/9912320}}.

\bibitem{Biswas:2010sa}
S.~Biswas, K.~Melnikov, and M.~Schulze, ``{Next-to-leading order QCD effects
  and the top quark mass measurements at the LHC},'' {\em JHEP} {\bf 1008}
  (2010) 048,
\href{http://xxx.lanl.gov/abs/1006.0910}{{\tt 1006.0910}}.

\bibitem{Frixione:2014ala}
S.~Frixione and A.~Mitov, ``{Determination of the top quark mass from leptonic
  observables},'' {\em JHEP} {\bf 1409} (2014) 012,
\href{http://xxx.lanl.gov/abs/1407.2763}{{\tt 1407.2763}}.

\bibitem{Chatrchyan:2013boa}
{\bf CMS} Collaboration, S.~Chatrchyan {\em et.~al.}, ``{Measurement of masses
  in the $t \bar{t}$ system by kinematic endpoints in pp collisions at
  $\sqrt{s}$ = 7 TeV},'' {\em Eur.Phys.J.} {\bf C73} (2013) 2494,
\href{http://xxx.lanl.gov/abs/1304.5783}{{\tt 1304.5783}}.

\bibitem{Frederix:2007gi}
R.~Frederix and F.~Maltoni, ``{Top pair invariant mass distribution: A Window
  on new physics},'' {\em JHEP} {\bf 0901} (2009) 047,
\href{http://xxx.lanl.gov/abs/0712.2355}{{\tt 0712.2355}}.

\bibitem{Alioli:2012hj}
S.~Alioli, J.~Fuster, A.~Irles, S.-O. Moch, P.~Uwer, {\em et.~al.},
  ``{Top-quark pair-production with one jet and parton showering at hadron
  colliders},''
\href{http://xxx.lanl.gov/abs/1206.1750}{{\tt 1206.1750}}.

\bibitem{Alioli:2013mxa}
S.~Alioli, P.~Fernandez, J.~Fuster, A.~Irles, S.-O. Moch, {\em et.~al.}, ``{A
  new observable to measure the top-quark mass at hadron colliders},''
\href{http://xxx.lanl.gov/abs/1303.6415}{{\tt 1303.6415}}.

\bibitem{Beneke:2011mq}
M.~Beneke, P.~Falgari, S.~Klein, and C.~Schwinn, ``{Hadronic top-quark pair
  production with NNLL threshold resummation},'' {\em Nucl.Phys.} {\bf B855}
  (2012) 695--741,
\href{http://xxx.lanl.gov/abs/1109.1536}{{\tt 1109.1536}}.

\bibitem{Ahrens:2011px}
V.~Ahrens, A.~Ferroglia, B.~D. Pecjak, and L.~L. Yang, ``{Precision predictions
  for the t+t(bar) production cross section at hadron colliders},'' {\em Phys.
  Lett.} {\bf B703} (2011) 135--141,
\href{http://xxx.lanl.gov/abs/1105.5824}{{\tt 1105.5824}}.

\bibitem{Giudice:2015toa}
G.~F. Giudice, P.~Paradisi, and A.~Strumia, ``{Indirect determinations of the
  top quark mass},''
\href{http://xxx.lanl.gov/abs/1508.05332}{{\tt 1508.05332}}.

\bibitem{Czarnecki:1990pe}
A.~Czarnecki, M.~Jezabek, and J.~H. Kuhn, ``{Lepton Spectra From Decays of
  Polarized Top Quarks},'' {\em Nucl. Phys.} {\bf B351} (1991)
70--80.

\bibitem{Brandenburg:2002xr}
A.~Brandenburg, Z.~G. Si, and P.~Uwer, ``{QCD corrected spin analyzing power of
  jets in decays of polarized top quarks},'' {\em Phys. Lett.} {\bf B539}
  (2002) 235--241,
\href{http://xxx.lanl.gov/abs/hep-ph/0205023}{{\tt hep-ph/0205023}}.

\bibitem{Bernreuther:2000yn}
W.~Bernreuther, A.~Brandenburg, and Z.~Si, ``{Next-to-leading order QCD
  corrections to top quark spin correlations at hadron colliders: The Reactions
  $q\bar{q} \rightarrow t \bar{t}(g)$},'' {\em Phys.Lett.} {\bf B483} (2000)
  99--104,
\href{http://xxx.lanl.gov/abs/hep-ph/0004184}{{\tt hep-ph/0004184}}.

\bibitem{Bernreuther:2001bx}
W.~Bernreuther, A.~Brandenburg, Z.~Si, and P.~Uwer, ``{Next-to-leading order
  QCD corrections to top quark spin correlations at hadron colliders: The
  Reactions $gg \rightarrow t \bar{t}(g)$ and $gq(\bar{q}) \rightarrow t
  \bar{t}q(\bar{q})$},'' {\em Phys.Lett.} {\bf B509} (2001) 53--58,
\href{http://xxx.lanl.gov/abs/hep-ph/0104096}{{\tt hep-ph/0104096}}.

\bibitem{Mahlon:1997uc}
G.~Mahlon and S.~Parke, ``Maximizing spin correlations in top quark pair
  production at the tevatron,'' {\em Phys. Lett.} {\bf B411} (1997) 173,
\href{http://xxx.lanl.gov/abs/hep-ph/9706304}{{\tt hep-ph/9706304}}.

\bibitem{Bernreuther:2004jv}
W.~Bernreuther, A.~Brandenburg, Z.~Si, and P.~Uwer, ``{Top quark pair
  production and decay at hadron colliders},'' {\em Nucl.Phys.} {\bf B690}
  (2004) 81--137,
\href{http://xxx.lanl.gov/abs/hep-ph/0403035}{{\tt hep-ph/0403035}}.

\bibitem{Mahlon:2010gw}
G.~Mahlon and S.~J. Parke, ``{Spin Correlation Effects in Top Quark Pair
  Production at the LHC},'' {\em Phys.Rev.} {\bf D81} (2010) 074024,
\href{http://xxx.lanl.gov/abs/1001.3422}{{\tt 1001.3422}}.

\bibitem{Melnikov:2011ai}
K.~Melnikov and M.~Schulze, ``{Top quark spin correlations at the Tevatron and
  the LHC},'' {\em Phys.Lett.} {\bf B700} (2011) 17--20,
\href{http://xxx.lanl.gov/abs/1103.2122}{{\tt 1103.2122}}.

\bibitem{Godbole:2010kr}
R.~M. Godbole, K.~Rao, S.~D. Rindani, and R.~K. Singh, ``{On measurement of top
  polarization as a probe of $t \bar t$ production mechanisms at the LHC},''
  {\em JHEP} {\bf 1011} (2010) 144,
\href{http://xxx.lanl.gov/abs/1010.1458}{{\tt 1010.1458}}.

\bibitem{Choudhury:2010cd}
D.~Choudhury, R.~M. Godbole, S.~D. Rindani, and P.~Saha, ``{Top polarization,
  forward-backward asymmetry and new physics},'' {\em Phys.Rev.} {\bf D84}
  (2011) 014023,
\href{http://xxx.lanl.gov/abs/1012.4750}{{\tt 1012.4750}}.

\bibitem{Huitu:2010ad}
K.~Huitu, S.~Kumar~Rai, K.~Rao, S.~D. Rindani, and P.~Sharma, ``{Probing top
  charged-Higgs production using top polarization at the Large Hadron
  Collider},'' {\em JHEP} {\bf 04} (2011) 026,
\href{http://xxx.lanl.gov/abs/1012.0527}{{\tt 1012.0527}}.

\bibitem{Godbole:2011vw}
R.~M. Godbole, L.~Hartgring, I.~Niessen, and C.~D. White, ``{Top polarisation
  studies in $H^-t$ and $Wt$ production},'' {\em JHEP} {\bf 1201} (2012) 011,
\href{http://xxx.lanl.gov/abs/1111.0759}{{\tt 1111.0759}}.

\bibitem{Aaltonen:2013wca}
{\bf CDF Collaboration, D0} Collaboration, T.~A. Aaltonen {\em et.~al.},
  ``{Combination of measurements of the top-quark pair production cross section
  from the Tevatron Collider},'' {\em Phys.Rev.} {\bf D89} (2014) 072001,
\href{http://xxx.lanl.gov/abs/1309.7570}{{\tt 1309.7570}}.

\bibitem{Czakon:2013goa}
M.~Czakon, P.~Fiedler, and A.~Mitov, ``{The total top quark pair production
  cross-section at hadron colliders through ${\cal O}(\alpha_S^4)$},'' {\em
  Phys.Rev.Lett.} {\bf 110} (2013) 252004,
\href{http://xxx.lanl.gov/abs/1303.6254}{{\tt 1303.6254}}.

\bibitem{ATLAS:2012dpa}
{\bf ATLAS} Collaboration,
``{Combination of ATLAS and CMS top-quark pair cross section measurements using
  up to 1.1 fb-1 of data at 7 TeV},''.

\bibitem{CMS:2013sca}
{\bf CMS} Collaboration, C.~Collaboration,
``{Combination of ATLAS and CMS top-quark pair cross section measurements using
  proton-proton collisions at sqrt(s) = 7 TeV},''.

\bibitem{ATLAS:2014xxx}
T.~A. collaboration,
``{Combination of ATLAS and CMS top-quark pair cross-section measurements using
  proton-proton collisions at $\sqrt{s} = 8$ TeV},''.

\bibitem{CMS:2014gta}
{\bf CMS} Collaboration, C.~Collaboration,
``{Combination of ATLAS and CMS top quark pair cross section measurements in
  the emu final state using proton-proton collisions at 8 TeV},''.

\bibitem{Khachatryan:2015uqb}
{\bf CMS} Collaboration, V.~Khachatryan {\em et.~al.}, ``{Measurement of the
  top quark pair production cross section in proton-proton collisions at
  sqrt(s)=13 TeV},''
\href{http://xxx.lanl.gov/abs/1510.05302}{{\tt 1510.05302}}.

\bibitem{Nason:1987xz}
P.~Nason, S.~Dawson, and R.~K. Ellis, ``{The Total Cross-Section for the
  Production of Heavy Quarks in Hadronic Collisions},'' {\em Nucl. Phys.} {\bf
  B303} (1988)
607.

\bibitem{Nason:1989zy}
P.~Nason, S.~Dawson, and R.~K. Ellis, ``The one particle inclusive differential
  cross-section for heavy quark production in hadronic collisions,'' {\em Nucl.
  Phys.} {\bf B327} (1989)
49--92.

\bibitem{Beenakker:1989bq}
W.~Beenakker, H.~Kuijf, W.~L. van Neerven, and J.~Smith, ``Qcd corrections to
  heavy quark production in p anti-p collisions,'' {\em Phys. Rev.} {\bf D40}
  (1989)
54--82.

\bibitem{Beenakker:1991ma}
W.~Beenakker, W.~L. van Neerven, R.~Meng, G.~A. Schuler, and J.~Smith, ``Qcd
  corrections to heavy quark production in hadron-hadron collisions,'' {\em
  Nucl. Phys.} {\bf B351} (1991)
507--560.

\bibitem{Mangano:1991jk}
M.~L. Mangano, P.~Nason, and G.~Ridolfi, ``Heavy quark correlations in hadron
  collisions at next-to- leading order,'' {\em Nucl. Phys.} {\bf B373} (1992)
295--345.

\bibitem{Frixione:2002ik}
S.~Frixione and B.~R. Webber, ``Matching nlo qcd computations and parton shower
  simulations,'' {\em JHEP} {\bf 06} (2002) 029,
\href{http://xxx.lanl.gov/abs/hep-ph/0204244}{{\tt hep-ph/0204244}}.

\bibitem{Frixione:2003ei}
S.~Frixione, P.~Nason, and B.~R. Webber, ``Matching nlo qcd and parton showers
  in heavy flavour production,'' {\em JHEP} {\bf 08} (2003) 007,
\href{http://xxx.lanl.gov/abs/hep-ph/0305252}{{\tt hep-ph/0305252}}.

\bibitem{Nason:2004rx}
P.~Nason, ``{A New method for combining NLO QCD with shower Monte Carlo
  algorithms},'' {\em JHEP} {\bf 0411} (2004) 040,
\href{http://xxx.lanl.gov/abs/hep-ph/0409146}{{\tt hep-ph/0409146}}.

\bibitem{Frixione:2007vw}
S.~Frixione, P.~Nason, and C.~Oleari, ``{Matching NLO QCD computations with
  Parton Shower simulations: the POWHEG method},'' {\em JHEP} {\bf 0711} (2007)
  070,
\href{http://xxx.lanl.gov/abs/0709.2092}{{\tt 0709.2092}}.

\bibitem{Baernreuther:2012ws}
P.~Baernreuther, M.~Czakon, and A.~Mitov, ``{Percent Level Precision Physics at
  the Tevatron: First Genuine NNLO QCD Corrections to $q \bar{q} \to t \bar{t}
  + X$},'' {\em Phys.Rev.Lett.} {\bf 109} (2012) 132001,
\href{http://xxx.lanl.gov/abs/1204.5201}{{\tt 1204.5201}}.

\bibitem{Czakon:2012zr}
M.~Czakon and A.~Mitov, ``{NNLO corrections to top-pair production at hadron
  colliders: the all-fermionic scattering channels},'' {\em JHEP} {\bf 1212}
  (2012) 054,
\href{http://xxx.lanl.gov/abs/1207.0236}{{\tt 1207.0236}}.

\bibitem{Czakon:2012pz}
M.~Czakon and A.~Mitov, ``{NNLO corrections to top pair production at hadron
  colliders: the quark-gluon reaction},'' {\em JHEP} {\bf 1301} (2013) 080,
\href{http://xxx.lanl.gov/abs/1210.6832}{{\tt 1210.6832}}.

\bibitem{Czakon:2011ve}
M.~Czakon, ``{Double-real radiation in hadronic top quark pair production as a
  proof of a certain concept},'' {\em Nucl. Phys.} {\bf B849} (2011) 250--295,
\href{http://xxx.lanl.gov/abs/1101.0642}{{\tt 1101.0642}}.

\bibitem{Abelof:2011jv}
G.~Abelof and A.~Gehrmann-De~Ridder, ``{Antenna subtraction for the production
  of heavy particles at hadron colliders},'' {\em JHEP} {\bf 1104} (2011) 063,
\href{http://xxx.lanl.gov/abs/1102.2443}{{\tt 1102.2443}}.

\bibitem{Bernreuther:2011jt}
W.~Bernreuther, C.~Bogner, and O.~Dekkers, ``{The real radiation antenna
  function for $S \to Q {\bar Q} q {\bar q}$ at NNLO QCD},'' {\em JHEP} {\bf
  1106} (2011) 032,
\href{http://xxx.lanl.gov/abs/1105.0530}{{\tt 1105.0530}}.

\bibitem{Dittmaier:2007wz}
S.~Dittmaier, P.~Uwer, and S.~Weinzierl, ``{NLO QCD corrections to t anti-t +
  jet production at hadron colliders},'' {\em Phys. Rev. Lett.} {\bf 98} (2007)
  262002,
\href{http://xxx.lanl.gov/abs/hep-ph/0703120}{{\tt hep-ph/0703120}}.

\bibitem{Melnikov:2010iu}
K.~Melnikov and M.~Schulze, ``{NLO QCD corrections to top quark pair production
  in association with one hard jet at hadron colliders},'' {\em Nucl.Phys.}
  {\bf B840} (2010) 129--159,
\href{http://xxx.lanl.gov/abs/1004.3284}{{\tt 1004.3284}}.

\bibitem{Czakon:2007ej}
M.~Czakon, A.~Mitov, and S.~Moch, ``{Heavy-quark production in massless quark
  scattering at two loops in QCD},'' {\em Phys.Lett.} {\bf B651} (2007)
  147--159,
\href{http://xxx.lanl.gov/abs/0705.1975}{{\tt 0705.1975}}.

\bibitem{Czakon:2007wk}
M.~Czakon, A.~Mitov, and S.~Moch, ``{Heavy-quark production in gluon fusion at
  two loops in QCD},'' {\em Nucl.Phys.} {\bf B798} (2008) 210--250,
\href{http://xxx.lanl.gov/abs/0707.4139}{{\tt 0707.4139}}.

\bibitem{Bernreuther:2004ih}
W.~Bernreuther, R.~Bonciani, T.~Gehrmann, R.~Heinesch, T.~Leineweber, {\em
  et.~al.}, ``{Two-loop QCD corrections to the heavy quark form-factors: The
  Vector contributions},'' {\em Nucl.Phys.} {\bf B706} (2005) 245--324,
\href{http://xxx.lanl.gov/abs/hep-ph/0406046}{{\tt hep-ph/0406046}}.

\bibitem{Bonciani:2010mn}
R.~Bonciani, A.~Ferroglia, T.~Gehrmann, A.~Manteuffel, and C.~Studerus,
  ``{Two-Loop Leading Color Corrections to Heavy-Quark Pair Production in the
  Gluon Fusion Channel},'' {\em JHEP} {\bf 1101} (2011) 102,
\href{http://xxx.lanl.gov/abs/1011.6661}{{\tt 1011.6661}}.

\bibitem{Baernreuther:2013caa}
P.~BŠrnreuther, M.~Czakon, and P.~Fiedler, ``{Virtual amplitudes and threshold
  behaviour of hadronic top-quark pair-production cross sections},'' {\em JHEP}
  {\bf 02} (2014) 078,
\href{http://xxx.lanl.gov/abs/1312.6279}{{\tt 1312.6279}}.

\bibitem{Czakon:2013tha}
M.~Czakon, M.~L. Mangano, A.~Mitov, and J.~Rojo, ``{Constraints on the gluon
  PDF from top quark pair production at hadron colliders},''
\href{http://xxx.lanl.gov/abs/1303.7215}{{\tt 1303.7215}}.

\bibitem{Ball:2014uwa}
{\bf NNPDF} Collaboration, R.~D. Ball {\em et.~al.}, ``{Parton distributions
  for the LHC Run II},'' {\em JHEP} {\bf 04} (2015) 040,
\href{http://xxx.lanl.gov/abs/1410.8849}{{\tt 1410.8849}}.

\bibitem{Bernreuther:2006vg}
W.~Bernreuther, M.~Fuecker, and Z.-G. Si, ``{Weak interaction corrections to
  hadronic top quark pair production},'' {\em Phys. Rev.} {\bf D74} (2006)
  113005,
\href{http://xxx.lanl.gov/abs/hep-ph/0610334}{{\tt hep-ph/0610334}}.

\bibitem{Kuhn:2006vh}
J.~H. Kuhn, A.~Scharf, and P.~Uwer, ``{Electroweak effects in top-quark pair
  production at hadron colliders},'' {\em Eur.Phys.J.} {\bf C51} (2007) 37--53,
\href{http://xxx.lanl.gov/abs/hep-ph/0610335}{{\tt hep-ph/0610335}}.

\bibitem{Moretti:2006nf}
S.~Moretti, M.~Nolten, and D.~Ross, ``{Weak corrections to gluon-induced
  top-antitop hadro-production},'' {\em Phys.Lett.} {\bf B639} (2006) 513--519,
\href{http://xxx.lanl.gov/abs/hep-ph/0603083}{{\tt hep-ph/0603083}}.

\bibitem{Kuhn:2011ri}
J.~H. Kuhn and G.~Rodrigo, ``{Charge asymmetries of top quarks at hadron
  colliders revisited},'' {\em JHEP} {\bf 1201} (2012) 063,
\href{http://xxx.lanl.gov/abs/1109.6830}{{\tt 1109.6830}}.

\bibitem{Bevilacqua:2010qb}
G.~Bevilacqua, M.~Czakon, A.~van Hameren, C.~G. Papadopoulos, and M.~Worek,
  ``{Complete off-shell effects in top quark pair hadroproduction with leptonic
  decay at next-to-leading order},'' {\em JHEP} {\bf 1102} (2011) 083,
\href{http://xxx.lanl.gov/abs/1012.4230}{{\tt 1012.4230}}.

\bibitem{Kidonakis:2001nj}
N.~Kidonakis, E.~Laenen, S.~Moch, and R.~Vogt, ``{Sudakov resummation and
  finite order expansions of heavy quark hadroproduction cross sections},''
  {\em Phys. Rev.} {\bf D64} (2001) 114001,
\href{http://xxx.lanl.gov/abs/hep-ph/0105041}{{\tt hep-ph/0105041}}.

\bibitem{Sterman:2000pt}
G.~Sterman and W.~Vogelsang, ``Threshold resummation and rapidity dependence,''
  {\em JHEP} {\bf 02} (2001) 016,
\href{http://xxx.lanl.gov/abs/http://arXiv.org/abs/hep-ph/0011289}{{\tt
  http://arXiv.org/abs/hep-ph/0011289}}.

\bibitem{Czakon:2009zw}
M.~Czakon, A.~Mitov, and G.~F. Sterman, ``{Threshold Resummation for Top-Pair
  Hadroproduction to Next-to-Next-to-Leading Log},'' {\em Phys.Rev.} {\bf D80}
  (2009) 074017,
\href{http://xxx.lanl.gov/abs/0907.1790}{{\tt 0907.1790}}.

\bibitem{Ahrens:2011mw}
V.~Ahrens, A.~Ferroglia, M.~Neubert, B.~D. Pecjak, and L.~L. Yang,
  ``{RG-improved single-particle inclusive cross sections and forward-backward
  asymmetry in $t\bar t$ production at hadron colliders},''
\href{http://xxx.lanl.gov/abs/1103.0550}{{\tt 1103.0550}}.

\bibitem{Kidonakis:2009ev}
N.~Kidonakis, ``{Two-loop soft anomalous dimensions and NNLL resummation for
  heavy quark production},'' {\em Phys. Rev. Lett.} {\bf 102} (2009) 232003,
\href{http://xxx.lanl.gov/abs/0903.2561}{{\tt 0903.2561}}.

\bibitem{Kidonakis:2010dk}
N.~Kidonakis, ``{Next-to-next-to-leading soft-gluon corrections for the top
  quark cross section and transverse momentum distribution},'' {\em Phys. Rev.}
  {\bf D82} (2010) 114030,
\href{http://xxx.lanl.gov/abs/1009.4935}{{\tt 1009.4935}}.

\bibitem{Kidonakis:2014isa}
N.~Kidonakis, ``{NNNLO soft-gluon corrections for the top-antitop pair
  production cross section},'' {\em Phys. Rev.} {\bf D90} (2014), no.~1,
  014006,
\href{http://xxx.lanl.gov/abs/1405.7046}{{\tt 1405.7046}}.

\bibitem{Aliev:2010zk}
M.~Aliev, H.~Lacker, U.~Langenfeld, S.~Moch, P.~Uwer, {\em et.~al.}, ``{HATHOR:
  HAdronic Top and Heavy quarks crOss section calculatoR},'' {\em
  Comput.Phys.Commun.} {\bf 182} (2011) 1034--1046,
\href{http://xxx.lanl.gov/abs/1007.1327}{{\tt 1007.1327}}.

\bibitem{Czakon:2011xx}
M.~Czakon and A.~Mitov, ``{Top++: A Program for the Calculation of the Top-Pair
  Cross-Section at Hadron Colliders},''
\href{http://xxx.lanl.gov/abs/1112.5675}{{\tt 1112.5675}}.

\bibitem{Beneke:2012wb}
M.~Beneke, P.~Falgari, S.~Klein, J.~Piclum, C.~Schwinn, {\em et.~al.},
  ``{Inclusive Top-Pair Production Phenomenology with TOPIXS},'' {\em JHEP}
  {\bf 1207} (2012) 194,
\href{http://xxx.lanl.gov/abs/1206.2454}{{\tt 1206.2454}}.

\bibitem{Denner:2010jp}
A.~Denner, S.~Dittmaier, S.~Kallweit, and S.~Pozzorini, ``{NLO QCD corrections
  to WWbb production at hadron colliders},'' {\em Phys. Rev. Lett.} {\bf 106}
  (2011) 052001,
\href{http://xxx.lanl.gov/abs/1012.3975}{{\tt 1012.3975}}.

\bibitem{Kuhn:1998kw}
J.~H. Kuhn and G.~Rodrigo, ``{Charge asymmetry of heavy quarks at hadron
  colliders},'' {\em Phys.Rev.} {\bf D59} (1999) 054017,
  \href{http://xxx.lanl.gov/abs/hep-ph/9807420}{{\tt hep-ph/9807420}}.

\bibitem{Aaltonen:2008hc}
{\bf CDF} Collaboration, T.~Aaltonen {\em et.~al.}, ``{Forward-Backward
  Asymmetry in Top Quark Production in $p\bar{p}$ Collisions at $sqrt{s}=1.96$
  TeV},'' {\em Phys.Rev.Lett.} {\bf 101} (2008) 202001,
  \href{http://xxx.lanl.gov/abs/0806.2472}{{\tt 0806.2472}}.

\bibitem{Aaltonen:2011kc}
{\bf CDF} Collaboration, T.~Aaltonen {\em et.~al.}, ``{Evidence for a Mass
  Dependent Forward-Backward Asymmetry in Top Quark Pair Production},'' {\em
  Phys. Rev.} {\bf D83} (2011) 112003,
\href{http://xxx.lanl.gov/abs/1101.0034}{{\tt 1101.0034}}.

\bibitem{Abazov:2007qb}
{\bf D0} Collaboration, V.~Abazov {\em et.~al.}, ``{First measurement of the
  forward-backward charge asymmetry in top quark pair production},'' {\em
  Phys.Rev.Lett.} {\bf 100} (2008) 142002,
  \href{http://xxx.lanl.gov/abs/0712.0851}{{\tt 0712.0851}}.

\bibitem{Abazov:2011rq}
{\bf D0} Collaboration, V.~M. Abazov {\em et.~al.}, ``{Forward-backward
  asymmetry in top quark-antiquark production},'' {\em Phys.Rev.} {\bf D84}
  (2011) 112005,
\href{http://xxx.lanl.gov/abs/1107.4995}{{\tt 1107.4995}}.

\bibitem{Kidonakis:1996aq}
N.~Kidonakis and G.~Sterman, ``Subleading logarithms in qcd hard scattering,''
  {\em Phys. Lett.} {\bf B387} (1996)
867--874.

\bibitem{Almeida:2008ug}
L.~G. Almeida, G.~F. Sterman, and W.~Vogelsang, ``{Threshold Resummation for
  the Top Quark Charge Asymmetry},'' {\em Phys.Rev.} {\bf D78} (2008) 014008,
  \href{http://xxx.lanl.gov/abs/0805.1885}{{\tt 0805.1885}}.

\bibitem{Ahrens:2011uf}
V.~Ahrens, A.~Ferroglia, M.~Neubert, B.~D. Pecjak, and L.~L. Yang, ``{The
  top-pair forward-backward asymmetry beyond NLO},'' {\em Phys.Rev.} {\bf D84}
  (2011) 074004,
\href{http://xxx.lanl.gov/abs/1106.6051}{{\tt 1106.6051}}.

\bibitem{Kidonakis:2011zn}
N.~Kidonakis, ``{The top quark rapidity distribution and forward-backward
  asymmetry},'' {\em Phys. Rev.} {\bf D84} (2011) 011504,
\href{http://xxx.lanl.gov/abs/1105.5167}{{\tt 1105.5167}}.

\bibitem{Kidonakis:2015ona}
N.~Kidonakis, ``{The top quark forward-backward asymmetry at approximate
  N$^3$LO},'' {\em Phys. Rev.} {\bf D91} (2015), no.~7, 071502,
\href{http://xxx.lanl.gov/abs/1501.01581}{{\tt 1501.01581}}.

\bibitem{Abazov:2013wxa}
{\bf D0} Collaboration, V.~M. Abazov {\em et.~al.}, ``{Measurement of the
  asymmetry in angular distributions of leptons produced in dilepton ttbar
  final states in ppbar collisions at sqrt(s)=1.96 TeV},'' {\em Phys.Rev.} {\bf
  D88} (2013) 112002,
\href{http://xxx.lanl.gov/abs/1308.6690}{{\tt 1308.6690}}.

\bibitem{Chapelain:2013eea}
{\bf D0} Collaboration, A.~Chapelain, ``{Measurement of the leptonic ttbar
  charge asymmetry in the dilepton channel with the D0 detector},''
\href{http://xxx.lanl.gov/abs/1311.6731}{{\tt 1311.6731}}.

\bibitem{Bernreuther:2012sx}
W.~Bernreuther and Z.-G. Si, ``{Top quark and leptonic charge asymmetries for
  the Tevatron and LHC},'' {\em Phys.Rev.} {\bf D86} (2012) 034026,
\href{http://xxx.lanl.gov/abs/1205.6580}{{\tt 1205.6580}}.

\bibitem{Hollik:2011ps}
W.~Hollik and D.~Pagani, ``{The electroweak contribution to the top quark
  forward-backward asymmetry at the Tevatron},'' {\em Phys.Rev.} {\bf D84}
  (2011) 093003,
\href{http://xxx.lanl.gov/abs/1107.2606}{{\tt 1107.2606}}.

\bibitem{Skands:2012mm}
P.~Skands, B.~Webber, and J.~Winter, ``{QCD Coherence and the Top Quark
  Asymmetry},'' {\em JHEP} {\bf 1207} (2012) 151,
\href{http://xxx.lanl.gov/abs/1205.1466}{{\tt 1205.1466}}.

\bibitem{Czakon:2014xsa}
M.~Czakon, P.~Fiedler, and A.~Mitov, ``{Resolving the Tevatron Top Quark
  Forward-Backward Asymmetry Puzzle: Fully Differential
  Next-to-Next-to-Leading-Order Calculation},'' {\em Phys. Rev. Lett.} {\bf
  115} (2015), no.~5, 052001,
\href{http://xxx.lanl.gov/abs/1411.3007}{{\tt 1411.3007}}.

\bibitem{Abazov:2014cca}
{\bf D0} Collaboration, V.~M. Abazov {\em et.~al.}, ``{Measurement of the
  forward-backward asymmetry in top quark-antiquark production in ppbar
  collisions using the lepton+jets channel},'' {\em Phys.Rev.} {\bf D90}
  (2014), no.~7, 072011,
\href{http://xxx.lanl.gov/abs/1405.0421}{{\tt 1405.0421}}.

\bibitem{Aaltonen:2012it}
{\bf CDF} Collaboration, T.~Aaltonen {\em et.~al.}, ``{Measurement of the top
  quark forward-backward production asymmetry and its dependence on event
  kinematic properties},'' {\em Phys.Rev.} {\bf D87} (2013) 092002,
\href{http://xxx.lanl.gov/abs/1211.1003}{{\tt 1211.1003}}.

\bibitem{Sharyy:2013dfa}
{\bf D0, CDF, ATLAS and CMS} Collaboration, V.~Sharyy, ``{Experimental status
  of top charge asymmetry measurements},''
\href{http://xxx.lanl.gov/abs/1312.0383}{{\tt 1312.0383}}.

\bibitem{Bevilacqua:2011aa}
G.~Bevilacqua, M.~Czakon, C.~Papadopoulos, and M.~Worek, ``{Hadronic top-quark
  pair production in association with two jets at Next-to-Leading Order QCD},''
  {\em Phys.Rev.} {\bf D84} (2011) 114017,
\href{http://xxx.lanl.gov/abs/1108.2851}{{\tt 1108.2851}}.

\bibitem{Hewett:2011wz}
J.~L. Hewett, J.~Shelton, M.~Spannowsky, T.~M. Tait, and M.~Takeuchi,
  ``{$A^t_{FB}$ Meets LHC},'' {\em Phys.Rev.} {\bf D84} (2011) 054005,
\href{http://xxx.lanl.gov/abs/1103.4618}{{\tt 1103.4618}}.

\bibitem{Berge:2012rc}
S.~Berge and S.~Westhoff, ``{Top-Quark Charge Asymmetry with a Jet Handle},''
  {\em Phys.Rev.} {\bf D86} (2012) 094036,
\href{http://xxx.lanl.gov/abs/1208.4104}{{\tt 1208.4104}}.

\bibitem{Berge:2013xsa}
S.~Berge and S.~Westhoff, ``{Top-Quark Charge Asymmetry Goes Forward: Two New
  Observables for Hadron Colliders},'' {\em JHEP} {\bf 1307} (2013) 179,
\href{http://xxx.lanl.gov/abs/1305.3272}{{\tt 1305.3272}}.

\bibitem{ATLASchargeasycomb}
{\bf ATLAS} Collaboration, ATLAS-CONF-2014-012, ``{Combination of ATLAS and CMS
  ttbar charge asymmetry measurements using LHC proton-proton collisions at 7
  TeV},''.

\bibitem{CMSchargeasycomb}
{\bf CMS} Collaboration, CMS-PAS-TOP-14-006, ``{Combination of ATLAS and CMS
  ttbar charge asymmetry measurements using LHC proton-proton collisions at 7
  TeV},''.

\bibitem{Ferroglia:2013zwa}
A.~Ferroglia, B.~D. Pecjak, and L.~L. Yang, ``{Top-quark pair production at
  high invariant mass: an NNLO soft plus virtual approximation},'' {\em JHEP}
  {\bf 1309} (2013) 032,
\href{http://xxx.lanl.gov/abs/1306.1537}{{\tt 1306.1537}}.

\bibitem{Ferroglia:2013awa}
A.~Ferroglia, S.~Marzani, B.~D. Pecjak, and L.~L. Yang, ``{Boosted top
  production: factorization and resummation for single-particle inclusive
  distributions},'' {\em JHEP} {\bf 01} (2014) 028,
\href{http://xxx.lanl.gov/abs/1310.3836}{{\tt 1310.3836}}.

\bibitem{Kidonakis:2014pja}
N.~Kidonakis, ``{NNNLO soft-gluon corrections for the top-quark $p_T$ and
  rapidity distributions},'' {\em Phys. Rev.} {\bf D91} (2015), no.~3, 031501,
\href{http://xxx.lanl.gov/abs/1411.2633}{{\tt 1411.2633}}.

\bibitem{Aad:2015eia}
{\bf ATLAS} Collaboration, G.~Aad {\em et.~al.}, ``{Differential top-antitop
  cross-section measurements as a function of observables constructed from
  final-state particles using pp collisions at $\sqrt{s}=7$ TeV in the ATLAS
  detector},'' {\em JHEP} {\bf 06} (2015) 100,
\href{http://xxx.lanl.gov/abs/1502.05923}{{\tt 1502.05923}}.

\bibitem{Khachatryan:2015oqa}
{\bf CMS} Collaboration, V.~Khachatryan {\em et.~al.}, ``{Measurement of the
  Differential Cross Section for Top Quark Pair Production in pp Collisions at
  $\sqrt{s}$ = 8 TeV},''
\href{http://xxx.lanl.gov/abs/1505.04480}{{\tt 1505.04480}}.

\bibitem{Chatrchyan:2012saa}
{\bf CMS} Collaboration, S.~Chatrchyan {\em et.~al.}, ``{Measurement of
  differential top-quark pair production cross sections in $pp$ colisions at
  $\sqrt{s}=7$ TeV},'' {\em Eur. Phys. J.} {\bf C73} (2013), no.~3, 2339,
\href{http://xxx.lanl.gov/abs/1211.2220}{{\tt 1211.2220}}.

\bibitem{CDF:2013gna}
{\bf CDF} Collaboration, T.~Aaltonen {\em et.~al.}, ``{Measurement of the
  Differential Cross Section $d{\sigma}/d(\cos {\theta}t)$ for Top-Quark Pair
  Production in $p-\bar{p}$ Collisions at $\sqrt{s} = 1.96$ TeV},'' {\em Phys.
  Rev. Lett.} {\bf 111} (2013), no.~18, 182002,
\href{http://xxx.lanl.gov/abs/1306.2357}{{\tt 1306.2357}}.

\bibitem{Jung:2012ih}
{\bf D0} Collaboration, A.~W. Jung, ``{Top differential cross section
  measurements (Tevatron)},'' {\em Nuovo Cim.} {\bf C035N3} (2012) 55--58,
\href{http://xxx.lanl.gov/abs/1201.2660}{{\tt 1201.2660}}.

\bibitem{Group:2009qk}
{\bf CDF Collaboration, D0} Collaboration, FERMILAB-CONF-14-370-E, ``{Tevatron
  measurement of single top quark cross sections and the CKM matrix element
  |Vtb|},''.

\bibitem{CDF:2014uma}
{\bf CDF Collaboration, D0} Collaboration, T.~A. Aaltonen {\em et.~al.},
  ``{Observation of s-channel production of single top quarks at the
  Tevatron},'' {\em Phys.Rev.Lett.} {\bf 112} (2014) 231803,
\href{http://xxx.lanl.gov/abs/1402.5126}{{\tt 1402.5126}}.

\bibitem{Aad:2014fwa}
{\bf ATLAS} Collaboration, G.~Aad {\em et.~al.}, ``{Comprehensive measurements
  of $t$-channel single top-quark production cross sections at $\sqrt{s} = 7$
  TeV with the ATLAS detector},'' {\em Phys. Rev.} {\bf D90} (2014), no.~11,
  112006,
\href{http://xxx.lanl.gov/abs/1406.7844}{{\tt 1406.7844}}.

\bibitem{Chatrchyan:2012ep}
{\bf CMS} Collaboration, S.~Chatrchyan {\em et.~al.}, ``{Measurement of the
  single-top-quark $t$-channel cross section in $pp$ collisions at $\sqrt{s}=7$
  TeV},'' {\em JHEP} {\bf 1212} (2012) 035,
\href{http://xxx.lanl.gov/abs/1209.4533}{{\tt 1209.4533}}.

\bibitem{ATLAS2014007}
ATLAS-CONF-2014-007.

\bibitem{Khachatryan:2014iya}
{\bf CMS} Collaboration, V.~Khachatryan {\em et.~al.}, ``{Measurement of the
  t-channel single-top-quark production cross section and of the $\mid V_{tb}
  \mid$ CKM matrix element in pp collisions at $\sqrt{s}$= 8 TeV},'' {\em JHEP}
  {\bf 1406} (2014) 090,
\href{http://xxx.lanl.gov/abs/1403.7366}{{\tt 1403.7366}}.

\bibitem{ATLASCMSsigmatcomb}
{\bf ATLAS Collaboration, CMS} Collaboration, ATLAS-CONF-2013-098,
  ``{Combination of single-top Quark cross-section measurements in the
  $t$-channel at $\sqrt{s}=8$ TeV with the ATLAS and CMS experiments},''.

\bibitem{ATLAS2015047}
ATLAS-CONF-2015-047.

\bibitem{Bordes:1994ki}
G.~Bordes and B.~van Eijk, ``{Calculating QCD corrections to single top
  production in hadronic interactions},'' {\em Nucl.Phys.} {\bf B435} (1995)
23--58.

\bibitem{Pittau:1996rp}
R.~Pittau, ``{Final state QCD corrections to off-shell single top production in
  hadron collisions},'' {\em Phys.Lett.} {\bf B386} (1996) 397--402,
\href{http://xxx.lanl.gov/abs/hep-ph/9603265}{{\tt hep-ph/9603265}}.

\bibitem{Stelzer:1997ns}
T.~Stelzer, Z.~Sullivan, and S.~Willenbrock, ``{Single top quark production via
  $W$ - gluon fusion at next-to-leading order},'' {\em Phys.Rev.} {\bf D56}
  (1997) 5919--5927,
\href{http://xxx.lanl.gov/abs/hep-ph/9705398}{{\tt hep-ph/9705398}}.

\bibitem{Harris:2002md}
B.~W. Harris, E.~Laenen, L.~Phaf, Z.~Sullivan, and S.~Weinzierl, ``The fully
  differential single top quark cross section in next-to-leading order qcd,''
  {\em Phys. Rev.} {\bf D66} (2002) 054024,
\href{http://xxx.lanl.gov/abs/hep-ph/0207055}{{\tt hep-ph/0207055}}.

\bibitem{Cao:2004ap}
Q.-H. Cao, R.~Schwienhorst, and C.~P. Yuan, ``Next-to-leading order corrections
  to single top quark production and decay at tevatron. i: s-channel process,''
  {\em Phys. Rev.} {\bf D71} (2005) 054023,
\href{http://xxx.lanl.gov/abs/hep-ph/0409040}{{\tt hep-ph/0409040}}.

\bibitem{Cao:2005pq}
Q.-H. Cao, R.~Schwienhorst, J.~A. Benitez, R.~Brock, and C.~P. Yuan,
  ``Next-to-leading order corrections to single top quark production and decay
  at the tevatron. ii: t-channel process,''
\href{http://xxx.lanl.gov/abs/hep-ph/0504230}{{\tt hep-ph/0504230}}.

\bibitem{Campbell:2004ch}
J.~Campbell, R.~K. Ellis, and F.~Tramontano, ``Single top production and decay
  at next-to-leading order,'' {\em Phys. Rev.} {\bf D70} (2004) 094012,
\href{http://xxx.lanl.gov/abs/hep-ph/0408158}{{\tt hep-ph/0408158}}.

\bibitem{Campbell:2005bb}
J.~Campbell and F.~Tramontano, ``Next-to-leading order corrections to w t
  production and decay,'' {\em Nucl. Phys.} {\bf B726} (2005) 109--130,
\href{http://xxx.lanl.gov/abs/hep-ph/0506289}{{\tt hep-ph/0506289}}.

\bibitem{Beccaria:2008av}
M.~Beccaria, C.~Carloni~Calame, G.~Macorini, E.~Mirabella, F.~Piccinini, {\em
  et.~al.}, ``{A Complete one-loop calculation of electroweak supersymmetric
  effects in t-channel single top production at CERN LHC},'' {\em Phys.Rev.}
  {\bf D77} (2008) 113018,
\href{http://xxx.lanl.gov/abs/0802.1994}{{\tt 0802.1994}}.

\bibitem{Campbell:2009gj}
J.~M. Campbell, R.~Frederix, F.~Maltoni, and F.~Tramontano, ``{NLO predictions
  for t-channel production of single top and fourth generation quarks at hadron
  colliders},'' {\em JHEP} {\bf 0910} (2009) 042,
\href{http://xxx.lanl.gov/abs/0907.3933}{{\tt 0907.3933}}.

\bibitem{Heim:2009ku}
S.~Heim, Q.-H. Cao, R.~Schwienhorst, and C.-P. Yuan, ``{Next-to-leading order
  QCD corrections to s-channel single top quark production and decay at the
  LHC},'' {\em Phys.Rev.} {\bf D81} (2010) 034005,
\href{http://xxx.lanl.gov/abs/0911.0620}{{\tt 0911.0620}}.

\bibitem{Falgari:2010sf}
P.~Falgari, P.~Mellor, and A.~Signer, ``{Production-decay interferences at NLO
  in QCD for $t$-channel single-top production},'' {\em Phys.Rev.} {\bf D82}
  (2010) 054028,
\href{http://xxx.lanl.gov/abs/1007.0893}{{\tt 1007.0893}}.

\bibitem{Schwienhorst:2010je}
R.~Schwienhorst, C.-P. Yuan, C.~Mueller, and Q.-H. Cao, ``{Single top quark
  production and decay in the $t$-channel at next-to-leading order at the
  LHC},'' {\em Phys.Rev.} {\bf D83} (2011) 034019,
\href{http://xxx.lanl.gov/abs/1012.5132}{{\tt 1012.5132}}.

\bibitem{Kidonakis:2006bu}
N.~Kidonakis, ``{Single top production at the Tevatron: Threshold resummation
  and finite-order soft gluon corrections},'' {\em Phys.Rev.} {\bf D74} (2006)
  114012,
\href{http://xxx.lanl.gov/abs/hep-ph/0609287}{{\tt hep-ph/0609287}}.

\bibitem{Wang:2010ue}
J.~Wang, C.~S. Li, H.~X. Zhu, and J.~J. Zhang, ``{Factorization and resummation
  of t-channel single top quark production},''
\href{http://xxx.lanl.gov/abs/1010.4509}{{\tt 1010.4509}}.

\bibitem{Zhu:2010mr}
H.~X. Zhu, C.~S. Li, J.~Wang, and J.~J. Zhang, ``{Factorization and resummation
  of s-channel single top quark production},'' {\em JHEP} {\bf 1102} (2011)
  099,
\href{http://xxx.lanl.gov/abs/1006.0681}{{\tt 1006.0681}}.

\bibitem{Kidonakis:2011wy}
N.~Kidonakis, ``{Next-to-next-to-leading-order collinear and soft gluon
  corrections for t-channel single top quark production},'' {\em Phys.Rev.}
  {\bf D83} (2011) 091503,
\href{http://xxx.lanl.gov/abs/1103.2792}{{\tt 1103.2792}}.

\bibitem{Frixione:2005vw}
S.~Frixione, E.~Laenen, P.~Motylinski, and B.~R. Webber, ``{Single-top
  production in MC@NLO},'' {\em JHEP} {\bf 0603} (2006) 092,
\href{http://xxx.lanl.gov/abs/hep-ph/0512250}{{\tt hep-ph/0512250}}.

\bibitem{Alioli:2009je}
S.~Alioli, P.~Nason, C.~Oleari, and E.~Re, ``{NLO single-top production matched
  with shower in POWHEG: s- and t-channel contributions},'' {\em JHEP} {\bf
  0909} (2009) 111,
\href{http://xxx.lanl.gov/abs/0907.4076}{{\tt 0907.4076}}.

\bibitem{Frederix:2012dh}
R.~Frederix, E.~Re, and P.~Torrielli, ``{Single-top t-channel hadroproduction
  in the four-flavour scheme with POWHEG and aMC@NLO},'' {\em JHEP} {\bf 1209}
  (2012) 130,
\href{http://xxx.lanl.gov/abs/1207.5391}{{\tt 1207.5391}}.

\bibitem{Brucherseifer:2014ama}
M.~Brucherseifer, F.~Caola, and K.~Melnikov, ``{On the NNLO QCD corrections to
  single-top production at the LHC},'' {\em Phys.Lett.} {\bf B736} (2014)
  58--63,
\href{http://xxx.lanl.gov/abs/1404.7116}{{\tt 1404.7116}}.

\bibitem{Wang:2012dc}
J.~Wang, C.~S. Li, and H.~X. Zhu, ``{Resummation prediction on top quark
  transverse momentum distribution at large $p_T$},'' {\em Phys. Rev.} {\bf
  D87} (2013), no.~3, 034030,
\href{http://xxx.lanl.gov/abs/1210.7698}{{\tt 1210.7698}}.

\bibitem{Kidonakis:2015wva}
N.~Kidonakis, ``{Single-top transverse-momentum distributions at approximate
  NNLO},''
\href{http://xxx.lanl.gov/abs/1510.06361}{{\tt 1510.06361}}.

\bibitem{Tait:1999cf}
T.~M.~P. Tait, ``The t w- mode of single top production,'' {\em Phys. Rev.}
  {\bf D61} (2000) 034001,
\href{http://xxx.lanl.gov/abs/hep-ph/9909352}{{\tt hep-ph/9909352}}.

\bibitem{Belyaev:1998dn}
A.~S. Belyaev, E.~E. Boos, and L.~V. Dudko, ``Single top quark at future hadron
  colliders: Complete signal and background study,'' {\em Phys. Rev.} {\bf D59}
  (1999) 075001,
\href{http://xxx.lanl.gov/abs/hep-ph/9806332}{{\tt hep-ph/9806332}}.

\bibitem{Kersevan:2006fq}
B.~P. Kersevan and I.~Hinchliffe, ``{A Consistent prescription for the
  production involving massive quarks in hadron collisions},'' {\em JHEP} {\bf
  0609} (2006) 033,
\href{http://xxx.lanl.gov/abs/hep-ph/0603068}{{\tt hep-ph/0603068}}.

\bibitem{Zhu:2002uj}
S.~Zhu, ``Next-to-leading order qcd corrections to b g $\to$ t w- at the cern
  large hadron collider,'' {\em Phys. Lett.} {\bf B524} (2002)
283--288.

\bibitem{Cao:2008af}
Q.-H. Cao, ``{Demonstration of One Cutoff Phase Space Slicing Method:
  Next-to-Leading Order QCD Corrections to the tW Associated Production in
  Hadron Collision},''
\href{http://xxx.lanl.gov/abs/0801.1539}{{\tt 0801.1539}}.

\bibitem{Frixione:2008yi}
S.~Frixione, E.~Laenen, P.~Motylinski, B.~R. Webber, and C.~D. White,
  ``{Single-top hadroproduction in association with a W boson},'' {\em JHEP}
  {\bf 0807} (2008) 029,
\href{http://xxx.lanl.gov/abs/0805.3067}{{\tt 0805.3067}}.

\bibitem{Re:2010bp}
E.~Re, ``{Single-top Wt-channel production matched with parton showers using
  the POWHEG method},'' {\em Eur.Phys.J.} {\bf C71} (2011) 1547,
\href{http://xxx.lanl.gov/abs/1009.2450}{{\tt 1009.2450}}.

\bibitem{Aad:2012xca}
{\bf ATLAS} Collaboration, G.~Aad {\em et.~al.}, ``{Evidence for the associated
  production of a $W$ boson and a top quark in ATLAS at $\sqrt{s}=7$ TeV},''
  {\em Phys.Lett.} {\bf B716} (2012) 142--159,
\href{http://xxx.lanl.gov/abs/1205.5764}{{\tt 1205.5764}}.

\bibitem{Chatrchyan:2012zca}
{\bf CMS} Collaboration, S.~Chatrchyan {\em et.~al.}, ``{Evidence for
  associated production of a single top quark and W boson in pp collisions at 7
  TeV},'' {\em Phys.Rev.Lett.} (2012)
\href{http://xxx.lanl.gov/abs/1209.3489}{{\tt 1209.3489}}.

\bibitem{Aad:2015eto}
{\bf ATLAS} Collaboration, G.~Aad {\em et.~al.}, ``{Measurement of the
  production cross-section of a single top quark in association with a $W$
  boson at 8 TeV with the ATLAS experiment},''
\href{http://xxx.lanl.gov/abs/1510.03752}{{\tt 1510.03752}}.

\bibitem{CMSPASTOP14009}
CMS-PAS-TOP-14-009.

\bibitem{Kidonakis:2010ux}
N.~Kidonakis, ``{Two-loop soft anomalous dimensions for single top quark
  associated production with a W- or H-},'' {\em Phys.Rev.} {\bf D82} (2010)
  054018,
\href{http://xxx.lanl.gov/abs/1005.4451}{{\tt 1005.4451}}.

\bibitem{White:2009yt}
C.~D. White, S.~Frixione, E.~Laenen, and F.~Maltoni, ``{Isolating Wt production
  at the LHC},'' {\em JHEP} {\bf 0911} (2009) 074,
\href{http://xxx.lanl.gov/abs/0908.0631}{{\tt 0908.0631}}.

\bibitem{Frederix:2013gra}
R.~Frederix, ``{Top Quark Induced Backgrounds to Higgs Production in the
  $WW^{(*)}\to ll\nu\nu$ Decay Channel at Next-to-Leading-Order in QCD},'' {\em
  Phys.Rev.Lett.} {\bf 112} (2014), no.~8, 082002,
\href{http://xxx.lanl.gov/abs/1311.4893}{{\tt 1311.4893}}.

\bibitem{Cascioli:2013wga}
F.~Cascioli, S.~Kallweit, P.~Maierhöfer, and S.~Pozzorini, ``{A unified NLO
  description of top-pair and associated Wt production},'' {\em Eur.Phys.J.}
  {\bf C74} (2014), no.~3, 2783,
\href{http://xxx.lanl.gov/abs/1312.0546}{{\tt 1312.0546}}.

\bibitem{Beenakker:2002nc}
W.~Beenakker, S.~Dittmaier, M.~Kramer, B.~Plumper, M.~Spira, {\em et.~al.},
  ``{NLO QCD corrections to t anti-t H production in hadron collisions},'' {\em
  Nucl.Phys.} {\bf B653} (2003) 151--203,
\href{http://xxx.lanl.gov/abs/hep-ph/0211352}{{\tt hep-ph/0211352}}.

\bibitem{Dawson:2003zu}
S.~Dawson, C.~Jackson, L.~Orr, L.~Reina, and D.~Wackeroth, ``{Associated Higgs
  production with top quarks at the large hadron collider: NLO QCD
  corrections},'' {\em Phys.Rev.} {\bf D68} (2003) 034022,
\href{http://xxx.lanl.gov/abs/hep-ph/0305087}{{\tt hep-ph/0305087}}.

\bibitem{Frederix:2011zi}
R.~Frederix, S.~Frixione, V.~Hirschi, F.~Maltoni, R.~Pittau, {\em et.~al.},
  ``{Scalar and pseudoscalar Higgs production in association with a top-antitop
  pair},'' {\em Phys.Lett.} {\bf B701} (2011) 427--433,
\href{http://xxx.lanl.gov/abs/1104.5613}{{\tt 1104.5613}}.

\bibitem{Garzelli:2011vp}
M.~Garzelli, A.~Kardos, C.~Papadopoulos, and Z.~Trocsanyi, ``{Standard Model
  Higgs boson production in association with a top anti-top pair at NLO with
  parton showering},'' {\em Europhys.Lett.} {\bf 96} (2011) 11001,
\href{http://xxx.lanl.gov/abs/1108.0387}{{\tt 1108.0387}}.

\bibitem{Frixione:2014qaa}
S.~Frixione, V.~Hirschi, D.~Pagani, H.~Shao, and M.~Zaro, ``{Weak corrections
  to Higgs hadroproduction in association with a top-quark pair},'' {\em JHEP}
  {\bf 1409} (2014) 065,
\href{http://xxx.lanl.gov/abs/1407.0823}{{\tt 1407.0823}}.

\bibitem{Frixione:2015zaa}
S.~Frixione, V.~Hirschi, D.~Pagani, H.~S. Shao, and M.~Zaro, ``{Electroweak and
  QCD corrections to top-pair hadroproduction in association with heavy
  bosons},'' {\em JHEP} {\bf 06} (2015) 184,
\href{http://xxx.lanl.gov/abs/1504.03446}{{\tt 1504.03446}}.

\bibitem{Melnikov:2011qx}
K.~Melnikov, A.~Scharf, and M.~Schulze, ``{Top quark pair production in
  association with a jet: QCD corrections and jet radiation in top quark
  decays},'' {\em Phys.Rev.} {\bf D85} (2012) 054002,
\href{http://xxx.lanl.gov/abs/1111.4991}{{\tt 1111.4991}}.

\bibitem{Bevilacqua:2015qha}
G.~Bevilacqua, H.~B. Hartanto, M.~Kraus, and M.~Worek, ``{Top Quark Pair
  Production in Association with a Jet with NLO QCD Off-Shell Effects at the
  Large Hadron Collider},''
\href{http://xxx.lanl.gov/abs/1509.09242}{{\tt 1509.09242}}.

\bibitem{Kardos:2011qa}
A.~Kardos, C.~Papadopoulos, and Z.~Trocsanyi, ``{Top quark pair production in
  association with a jet with NLO parton showering},'' {\em Phys.Lett.} {\bf
  B705} (2011) 76--81,
\href{http://xxx.lanl.gov/abs/1101.2672}{{\tt 1101.2672}}.

\bibitem{Alioli:2011as}
S.~Alioli, S.-O. Moch, and P.~Uwer, ``{Hadronic top-quark pair-production with
  one jet and parton showering},'' {\em JHEP} {\bf 1201} (2012) 137,
\href{http://xxx.lanl.gov/abs/1110.5251}{{\tt 1110.5251}}.

\bibitem{Czakon:2015cla}
M.~Czakon, H.~B. Hartanto, M.~Kraus, and M.~Worek, ``{Matching the Nagy-Soper
  parton shower at next-to-leading order},'' {\em JHEP} {\bf 06} (2015) 033,
\href{http://xxx.lanl.gov/abs/1502.00925}{{\tt 1502.00925}}.

\bibitem{Bevilacqua:2010ve}
G.~Bevilacqua, M.~Czakon, C.~Papadopoulos, and M.~Worek, ``{Dominant QCD
  Backgrounds in Higgs Boson Analyses at the LHC: A Study of $pp \rightarrow t
  \bar{t}$ + 2 jets at Next-To-Leading Order},'' {\em Phys.Rev.Lett.} {\bf 104}
  (2010) 162002,
\href{http://xxx.lanl.gov/abs/1002.4009}{{\tt 1002.4009}}.

\bibitem{Hoeche:2014qda}
S.~Hoeche, F.~Krauss, P.~Maierhoefer, S.~Pozzorini, M.~Schonherr, {\em
  et.~al.}, ``{Next-to-leading order QCD predictions for top-quark pair
  production with up to two jets merged with a parton shower},''
\href{http://xxx.lanl.gov/abs/1402.6293}{{\tt 1402.6293}}.

\bibitem{Bredenstein:2009aj}
A.~Bredenstein, A.~Denner, S.~Dittmaier, and S.~Pozzorini, ``{NLO QCD
  corrections to $pp \rightarrow t\bar{t} b \bar{b} + X$ at the LHC},'' {\em
  Phys.Rev.Lett.} {\bf 103} (2009) 012002,
\href{http://xxx.lanl.gov/abs/0905.0110}{{\tt 0905.0110}}.

\bibitem{Bevilacqua:2009zn}
G.~Bevilacqua, M.~Czakon, C.~Papadopoulos, R.~Pittau, and M.~Worek, ``{Assault
  on the NLO Wishlist: $pp \rightarrow t \bar{t} b \bar{b}$},'' {\em JHEP} {\bf
  0909} (2009) 109,
\href{http://xxx.lanl.gov/abs/0907.4723}{{\tt 0907.4723}}.

\bibitem{Bredenstein:2010rs}
A.~Bredenstein, A.~Denner, S.~Dittmaier, and S.~Pozzorini, ``{NLO QCD
  Corrections to Top Anti-Top Bottom Anti-Bottom Production at the LHC: 2. full
  hadronic results},'' {\em JHEP} {\bf 1003} (2010) 021,
\href{http://xxx.lanl.gov/abs/1001.4006}{{\tt 1001.4006}}.

\bibitem{Kardos:2013vxa}
A.~Kardos and Z.~Tr—cs‡nyi, ``{Hadroproduction of t anti-t pair with a b anti-b
  pair using PowHel},'' {\em J. Phys.} {\bf G41} (2014) 075005,
\href{http://xxx.lanl.gov/abs/1303.6291}{{\tt 1303.6291}}.

\bibitem{Cascioli:2013era}
F.~Cascioli, P.~Maierhoefer, N.~Moretti, S.~Pozzorini, and F.~Siegert, ``{NLO
  matching for ttbb production with massive b-quarks},''
\href{http://xxx.lanl.gov/abs/1309.5912}{{\tt 1309.5912}}.

\bibitem{Garzelli:2014aba}
M.~V. Garzelli, A.~Kardos, and Z.~Tr—cs‡nyi, ``{Hadroproduction of
  $t\bar{t}b\bar{b}$ final states at LHC: predictions at NLO accuracy matched
  with Parton Shower},'' {\em JHEP} {\bf 03} (2015) 083,
\href{http://xxx.lanl.gov/abs/1408.0266}{{\tt 1408.0266}}.

\bibitem{Melnikov:2011ta}
K.~Melnikov, M.~Schulze, and A.~Scharf, ``{QCD corrections to top quark pair
  production in association with a photon at hadron colliders},'' {\em
  Phys.Rev.} {\bf D83} (2011) 074013,
\href{http://xxx.lanl.gov/abs/1102.1967}{{\tt 1102.1967}}.

\bibitem{Kardos:2014zba}
A.~Kardos and Z.~Tr—cs‡nyi, ``{Hadroproduction of t anti-t pair in association
  with an isolated photon at NLO accuracy matched with parton shower},'' {\em
  JHEP} {\bf 05} (2015) 090,
\href{http://xxx.lanl.gov/abs/1406.2324}{{\tt 1406.2324}}.

\bibitem{Kardos:2014pba}
A.~Kardos and Z.~Tr—cs‡nyi, ``{Hadroproduction of tÐanti-t pair with two
  isolated photons with PowHel},'' {\em Nucl. Phys.} {\bf B897} (2015)
  717--731,
\href{http://xxx.lanl.gov/abs/1408.0278}{{\tt 1408.0278}}.

\bibitem{Lazopoulos:2008de}
A.~Lazopoulos, T.~McElmurry, K.~Melnikov, and F.~Petriello, ``{Next-to-leading
  order QCD corrections to $t \bar{t} Z$ production at the LHC},'' {\em
  Phys.Lett.} {\bf B666} (2008) 62--65,
\href{http://xxx.lanl.gov/abs/0804.2220}{{\tt 0804.2220}}.

\bibitem{Kardos:2011na}
A.~Kardos, Z.~Trocsanyi, and C.~Papadopoulos, ``{Top quark pair production in
  association with a Z-boson at NLO accuracy},'' {\em Phys.Rev.} {\bf D85}
  (2012) 054015,
\href{http://xxx.lanl.gov/abs/1111.0610}{{\tt 1111.0610}}.

\bibitem{Garzelli:2011is}
M.~Garzelli, A.~Kardos, C.~Papadopoulos, and Z.~Trocsanyi, ``{Z0 - boson
  production in association with a top anti-top pair at NLO accuracy with
  parton shower effects},'' {\em Phys.Rev.} {\bf D85} (2012) 074022,
\href{http://xxx.lanl.gov/abs/1111.1444}{{\tt 1111.1444}}.

\bibitem{Garzelli:2012bn}
M.~Garzelli, A.~Kardos, C.~Papadopoulos, and Z.~Trocsanyi, ``{t $\bar{t}$
  $W^{+-}$ and t $\bar{t}$ Z Hadroproduction at NLO accuracy in QCD with Parton
  Shower and Hadronization effects},'' {\em JHEP} {\bf 1211} (2012) 056,
\href{http://xxx.lanl.gov/abs/1208.2665}{{\tt 1208.2665}}.

\bibitem{Rontsch:2015una}
R.~Roentsch and M.~Schulze, ``{Probing top-Z dipole moments at the LHC and
  ILC},'' {\em JHEP} {\bf 08} (2015) 044,
\href{http://xxx.lanl.gov/abs/1501.05939}{{\tt 1501.05939}}.

\bibitem{Rontsch:2014cca}
R.~Roentsch and M.~Schulze, ``{Constraining couplings of top quarks to the Z
  boson in $ t\overline{t} $ + Z production at the LHC},'' {\em JHEP} {\bf 07}
  (2014) 091, \href{http://xxx.lanl.gov/abs/1404.1005}{{\tt 1404.1005}}.
[Erratum: JHEP09,132(2015)].

\bibitem{Badger:2010mg}
S.~Badger, J.~M. Campbell, and R.~Ellis, ``{QCD corrections to the hadronic
  production of a heavy quark pair and a W-boson including decay
  correlations},'' {\em JHEP} {\bf 1103} (2011) 027,
\href{http://xxx.lanl.gov/abs/1011.6647}{{\tt 1011.6647}}.

\bibitem{Campbell:2012dh}
J.~M. Campbell and R.~K. Ellis, ``{$t \bar{t} W^{+-}$ production and decay at
  NLO},'' {\em JHEP} {\bf 1207} (2012) 052,
\href{http://xxx.lanl.gov/abs/1204.5678}{{\tt 1204.5678}}.

\bibitem{Maltoni:2014zpa}
F.~Maltoni, M.~Mangano, I.~Tsinikos, and M.~Zaro, ``{Top-quark charge asymmetry
  and polarization in $t\overline{t}W^ï½±$ production at the LHC},'' {\em
  Phys.Lett.} {\bf B736} (2014) 252--260,
\href{http://xxx.lanl.gov/abs/1406.3262}{{\tt 1406.3262}}.

\bibitem{Maltoni:2015ena}
F.~Maltoni, D.~Pagani, and I.~Tsinikos, ``{Associated production of a top-quark
  pair with vector bosons at NLO in QCD: impact on $t \bar{t} H$ searches at
  the LHC},''
\href{http://xxx.lanl.gov/abs/1507.05640}{{\tt 1507.05640}}.

\bibitem{Bevilacqua:2012em}
G.~Bevilacqua and M.~Worek, ``{Constraining BSM Physics at the LHC: Four top
  final states with NLO accuracy in perturbative QCD},'' {\em JHEP} {\bf 1207}
  (2012) 111,
\href{http://xxx.lanl.gov/abs/1206.3064}{{\tt 1206.3064}}.

\bibitem{Campbell:2013yla}
J.~Campbell, R.~K. Ellis, and R.~R嗜tsch, ``{Single top production in
  association with a Z boson at the LHC},'' {\em Phys.Rev.} {\bf D87} (2013),
  no.~11, 114006,
\href{http://xxx.lanl.gov/abs/1302.3856}{{\tt 1302.3856}}.

\end{thebibliography}\endgroup


\end{document}